\documentclass[hyper,aps,eqsecnum,superscriptaddress,12pt]{revtex4-1}
\usepackage{bbm}
\usepackage{mathrsfs}
\usepackage{amsmath,amssymb,bm,comment}
\usepackage{hyperref}
\usepackage{graphicx}
\usepackage{bm}
\usepackage{stmaryrd}
\usepackage{color}
\usepackage{fancyhdr}
\usepackage{slashed}

\begin{document}

\title{ Second-order   chiral kinetic theory from Wigner function approach}

\author{Shi-Zheng Yang}
\email{yangsz@ustc.edu.cn}
\affiliation{Department of Modern Physics, University of Science and Technology
of China, Hefei, Anhui 230026, China}
\affiliation{Institute of Frontier and Interdisciplinary Science,
Key Laboratory of Particle Physics and Particle Irradiation (MOE), Shandong University, Qingdao, Shandong 266237, China}

\author{Shu-Xiang Ma}
\email{shuxiangma@mail.sdu.edu.cn}
\affiliation{School of Space  Science and Technology, Shandong University, Weihai, Shandong 264209, China}

\author{Jian-Hua Gao}
\email{gaojh@sdu.edu.cn}
\affiliation{Shandong Provincial Key Laboratory of Nuclear Science, Nuclear Energy Technology and Comprehensive Utilization, Weihai Frontier Innovation Institute of Nuclear Technology, School of Nuclear Science, Energy and Power Engineering, Shandong University, Shandong 250061, China}
\affiliation{Weihai Research Institute of Industrial Technology of Shandong University, Weihai
264209, China}

\begin{abstract}
We present a systematic method to derive the chiral kinetic theory  from the Wigner equations based on quantum field theory order by order. We take special effort to derive the chiral kinetic theory in seven-dimensional phase space from the eight-dimensional chiral kinetic theory. We give the seven-dimensional chiral kinetic theory  up to second order of semiclassical expansion. We find some new second-order  contributions compared with other approaches.
\end{abstract}

\maketitle
\thispagestyle{fancy}
\renewcommand{\headrulewidth}{0pt}

\section{Introduction}
\label{sec:intro}

In relativistic heavy-ion collisions, it is possible to generate  strong magnetic
field \cite{Bzdak:2011yy,Deng:2012pc,Bloczynski:2012en} and orbital angular momentum or vorticity \cite{Liang:2004ph,Gao:2007bc,Becattini:2007sr,Csernai:2013bqa,Jiang:2016woz,Deng:2016gyh,Pang:2016igs},
which can lead to novel chiral effects such as the chiral magnetic effects (CME)  \cite{Vilenkin:1980fu,Kharzeev:2007jp,Fukushima:2008xe},
chiral vorticity effects (CVE)\cite{Vilenkin:1978hb,Kharzeev:2007tn,Erdmenger:2008rm,Banerjee:2008th},
and  chiral separate effects (CSE) \cite{Son:2004tq,Metlitski:2005pr} and so on.
The research on these various chiral effects in relativistic heavy-ion collisions has greatly stimulated the rapid progress on the chiral kinetic theory (CKT) \cite{Gao:2012ix,Stephanov:2012ki,Son:2012zy,Chen:2012ca,Manuel:2013zaa,Chen:2014cla,Chen:2015gta,Hidaka:2016yjf,Mueller:2017lzw,
Huang:2018wdl,Hidaka:2018ekt,Gao:2018wmr,Gao:2018jsi,Liu:2018xip}. The recent  reviews on the chiral transport  can be found in Refs.
\cite{Liu:2020ymh,Gao:2020vbh,Gao:2020pfu,Hidaka:2022dmn}.
The CKT provides a consistent theoretical framework to merge the chiral anomaly  into the kinetic theory by the Berry monopole in momentum space
and can be applicable   not only in relativistic heavy-ion   collisions in high energy nuclear physics  but also
in   condensed matter physics \cite{Gorbar:2016ygi,Gorbar:2016sey,Gorbar:2016vvg},
 astrophysics, and  cosmology\cite{Yamamoto:2020zrs,Kamada:2022nyt}.

The CME, CVE, and CSE are all first-order chiral effects in terms of the expansion of spacetime gradient and electromagnetic field, which is equivalent to semiclassical $\hbar $ expansion for
Abelian gauge field. The well-established  chiral kinetic equation (CKE) is also valid up to first order.
There have been already  some  effort   to  go beyond  first order, studying the second-order chiral effects \cite{Kharzeev:2011ds,Jimenez-Alba:2015bia,Satow:2014lia,Abbasi:2018zoc,
Banerjee:2012iz,Bhattacharyya:2013ida,Megias:2014mba,Gorbar:2017toh} and  deriving the second-order chiral kinetic theory \cite{Gorbar:2017cwv,Hayata:2020sqz,Mameda:2023ueq}.
 In Ref.\cite{Gorbar:2017cwv}, second-order chiral kinetic equation in seven-dimensional phase space was derived from  quantum mechanics by the  Hamiltonian for a single chiral fermion, including only second order  in electromagnetic field.   In Refs.\cite{Hayata:2020sqz,Mameda:2023ueq}, second-order chiral kinetic equation in eight-dimensional phase space was derived from  quantum field theory by
 the Wigner function for chiral fermions, including second order in both electromagnetic field and spacetime gradient.
In this paper, we will take special effort to derive second-order chiral kinetic equation in seven-dimensional phase space from quantum field theory within Wigner function framework \cite{Heinz:1983nx,Elze:1986qd,Vasak:1987um,Zhuang:1995pd}.
Actually, it is  not a trivial task at all to integrate the eight-dimensional CKE to seven-dimensional CKE in Wigner function approach.
For brevity, we will restrict ourselves to uniform  electromagnetic field.

In Sec.\ref{sec:Wigner}, we will give a brief review of the Wigner function formalism for chiral system to keep our present work
self-contained.  In Sec.\ref{sec:CKT-7d}, we   derive  the eight-dimensional chiral kinetic equation  up to second order of
semiclassical $\hbar$ expansion and  integrate out time-like component of momentum and obtain  the seven-dimensional CKE. Finally, we summarize our results
in Sec.\ref{sec:summary}.

We  use the metric convention $g^{\mu\nu}=\mathrm{diag}(1,-1,-1,-1)$ and  Levi-Civita  tensor convention $\epsilon^{0123}=1$. We use the natural unit with $\hbar = c =1$ unless
otherwise specified.

\section{Wigner function formalism}
\label{sec:Wigner}

The Wigner function $W(x,p)$ for Dirac field is $4\times4$ matrix in spinor space and the matrix element is defined as
the ensemble expectation value of the Wigner operator \cite{Heinz:1983nx,Elze:1986qd,Vasak:1987um},
\begin{equation}
\label{wigner}
W_{\alpha\beta}(x,p) = \int\frac{d^4 y}{(2\pi)^4}
e^{-ip\cdot y}\langle \bar\psi_\beta\left(x+\frac{y}{2}\right)
U\left(x+\frac{y}{2},x-\frac{y}{2}\right)
\psi_{\alpha}\left(x-\frac{y}{2}\right)\rangle,
\end{equation}
where $U$ denotes the gauge link along the straight line between $x-y/2$ and $x+y/2$,
\begin{equation}
\label{link}
U\left(x+\frac{y}{2},x-\frac{y}{2}\right)  \equiv \textrm{Exp} \left(-i \int_{x-y/2}^{x+y/2} dz^\mu A_\mu (z)\right).
\end{equation}
We will  restrict ourselves to  the chiral  and collisionless fermionic system
under  a constant  external electromagnetic field $F_{\mu\nu}$ in space and time, i.e.  $\partial^\lambda_x F^{\mu\nu}=0$, hence we have
removed the path ordering of the gauge link.  It should be noted that the charge carried by fermion has been absorbed into
the electromagnetic vector potential $A_\mu$.

The Wigner equation for chiral particle under constant electromagnetic field tensor is given by \cite{Vasak:1987um},
\begin{equation}
\label{eq-c}
\gamma_\mu ( p^\mu +\frac{i}{2}\nabla^\mu)  W(x,p)=0 ,
\end{equation}
where $\gamma^{\mu}$'s are Dirac matrices and
$\nabla^\mu \equiv \partial^\mu_x - {F^\mu}_\nu\partial^\nu_p$. The subscript or superscript $x$ and $p$ of the derivative denote the gradient for the coordinate and momentum, respectively.
We can decompose the Wigner function  in terms of 16
independent generators of the Clifford algebra,
\begin{eqnarray}
\label{decomposition}
W&=& \frac{1}{4}\left[\mathscr{F}+i\gamma^5 \mathscr{P}
+\gamma^\mu \mathscr{V}_\mu   +\gamma^5 \gamma^\mu \mathscr{A}_\mu
+\frac{1}{2}\sigma^{\mu\nu} \mathscr{S}_{\mu\nu}\right],
\end{eqnarray}
where we have suppressed all the arguments of the Wigner function for brevity. In the subsequent sections, we will
suppress or keep the arguments of the function alternately for convenience.
At chiral limit, it is more convenient to define the chiral basis
\begin{eqnarray}
\label{chibasis}
\mathscr{J}_s^\mu\equiv\frac{1}{2}\left(\mathscr{V}^\mu +s \mathscr{A}^\mu\right),
\end{eqnarray}
with $s=+1$ or $-1$ corresponding to right-handed or left-handed component, respectively.
Substituting  Eq. (\ref{decomposition}) together with Eq. (\ref{chibasis}) into  Eq.(\ref{eq-c}), we find that
 the right-handed  or left-handed component will totally  decouple from all  the other components,
\begin{eqnarray}
\label{Js-eq}
\nabla_\mu\mathscr{J}_s^\mu &=& 0,\\
\label{Js-c1}
p_\mu\mathscr{J}_s^\mu &=&0,\\
\label{Js-c2}
p^\mu \mathscr{J}_{s}^\nu - p^\nu \mathscr{J}_{s}^\mu &=&-\frac{s}{2}\hbar \epsilon^{\mu\nu\rho\sigma}\nabla_\rho \mathscr{J}_{s\sigma},
\end{eqnarray}
where we have recovered the dependence on Planck constant $\hbar$.
To simplify the Wigner equations, we can resort to semiclassical expansion  in $\hbar$
\begin{eqnarray}
\label{Js-expansion}
\mathscr{J}^\mu_s &=& \mathscr{J}^{(0)\mu}_s +\hbar  \mathscr{J}^{(1)\mu}_s +\hbar^2  \mathscr{J}^{(2)\mu}_s
 +... ...
\end{eqnarray}
In the following, we will  suppress the subscript ``$s$'' in the expressions  for brevity.
Substituting these expansions into the Wigner equations and requiring the equations hold order by order gives rise to
 zeroth-order equation
\begin{eqnarray}
\label{Js-cs-0a}
p_\mu\mathscr{J}^{(0)\mu} &=& 0,\\
\label{Js-ev-0}
\nabla_\mu\mathscr{J}^{(0)\mu} &=&0,\\
\label{Js-cs-0b}
p^\mu \mathscr{J}^{(0)\nu} - p^\nu \mathscr{J}^{(0)\mu}
&=& 0,
\end{eqnarray}
and the $k$-th order equations with $n\ge 1$
\begin{eqnarray}
\label{Js-cs-1a}
p_\mu\mathscr{J}^{(k)\mu} &=&0,\\
\label{Js-ev-1}
\nabla_\mu\mathscr{J}^{(k)\mu} &=& 0,\\
\label{Js-cs-1b}
p^\mu \mathscr{J}^{(k)\nu} - p^\nu \mathscr{J}^{(k)\mu}
 &=& -\frac{s}{2}\epsilon^{\mu\nu\rho\sigma}\nabla_\rho \mathscr{J}_{\sigma}^{(k-1)}.
\end{eqnarray}
We note that all the components of chiral Wigner function $\mathscr{J}^\mu$ are coupled with each other. In order to disentangle the Wigner equations and find
the independent components or equations, we will apply the method proposed in \cite{Gao:2018wmr} by introducing  a time-like constant  4-vector $n^\mu$ with normalization $n^2=1$.
Then we can  decompose any 4-vector $X^\mu$ as
\begin{equation}
X^\mu=X_n n^\mu + \bar X^\mu,
\end{equation}
where $X_n=X\cdot n$ and $\bar X^\mu = \Delta^{\mu\nu}X_\nu$ with $\Delta^{\mu\nu}=g^{\mu\nu}-n^\mu n^\nu$.
We can also  decompose the antisymmetric  tensors  $F^{\mu\nu}$  as
\begin{eqnarray}
F^{\mu\nu}&=&E^\mu n^\nu -E^\nu n^\mu +\epsilon^{\mu\nu\rho\sigma}n_\rho B_\sigma.
\end{eqnarray}
It is easy to verify the following relations
\begin{eqnarray}
E^\mu = F^{\mu\nu}n_\nu ,\  B^\mu=\epsilon^{\mu\nu\rho\sigma}n_\nu F_{\rho\sigma}/2.
\end{eqnarray}
With such decomposition, the derivative operator $\nabla_\mu$ can be expressed as
\begin{eqnarray}
\label{nabla-full}
\nabla^\mu &=& n^\mu \nabla_n + \bar{\nabla}^\mu, \\
\label{nabla-n}
\nabla_n &=& n^\mu \nabla_\mu = \partial^x_n + E^\mu \partial^p_\mu,\\
\label{nabla-bar}
\bar{\nabla}^\mu &=&\bar{\partial}^x_\mu -E^\mu \partial^p_n -\bar \epsilon^{\mu\rho\nu }B_\rho\partial^p_\nu,
\end{eqnarray}
where we have defined the space-like antisymmetric Levi-Civita tensor by
\begin{eqnarray}
\bar\epsilon^{\mu\rho\nu}=\epsilon^{\mu\sigma\rho\nu}n_\sigma.
\end{eqnarray}

\section{The CKT derived from Wigner function}
\label{sec:CKT-7d}

In this section, we display a systematic  method to formalize the CKT  by  disentangling the Wigner equations order by order.
In particular, we show how to organize the chiral Wigner function and  CKT in  seven-dimensional phase space into a unified form.
We will present the  CKT in  seven-dimensional phase space up to second order.

\subsection{Zeroth-order CKT}
We start from the zeroth-order Wigner equations. With the time-like vector $n^\mu $, we can decompose zeroth order equation (\ref{Js-cs-0b}) as
\begin{eqnarray}
\label{Js-cs-0b-n}
\bar p^\mu \mathscr{J}_{n}^{(0)} - p_n \bar{\mathscr{J}}^{(0)\mu}
&=& 0,\\
\label{Js-cs-0b-bar}
\bar p^\mu \bar{\mathscr{J}}^{(0)\nu} - \bar p^\nu \bar{\mathscr{J}}^{(0)\mu}
&=& 0.
\end{eqnarray}
Then the first equation (\ref{Js-cs-0b-n}) gives the relation
\begin{eqnarray}
\label{barJs-Jsn-0}
\bar{ \mathscr{J}}^{(0)\mu}
 &=&\frac{1}{p_n}\bar  p^\mu \mathscr{J}_{n}^{(0)}.
\end{eqnarray}
which implies that only the time-like component $\mathscr{J}_{n}^{(0)}$ is independent. The Wigner equation automatically expresses other componets
in terms of $\mathscr{J}_{n}^{(0)}$.
With this relation above,  the second equation (\ref{Js-cs-0b-bar}) holds automatically.
Substituting Eq. (\ref{barJs-Jsn-0}) into Eq.(\ref{Js-cs-0a}) gives rise to
\begin{eqnarray}
 p^2 \frac{\mathscr{J}_{n}^{(0)} }{p_n }  &=& 0,
\end{eqnarray}
which implies the on-shell condition with the general solution given by
\begin{eqnarray}
\mathscr{J}^{(0)}_{n}&=& p_n  \mathcal{J}^{(0)}_{n} \delta\left(p^2\right),
\end{eqnarray}
where the distribution function $\mathcal{J}^{(0)}_{n}$ is regular at $p^2=0$.
It follows that the space-like component of the  chiral Wigner function is given by
\begin{eqnarray}
\label{Js-mu-0}
\bar{\mathscr{J}}^{(0)\mu}
 &=& \bar p^\mu\mathcal{J}^{(0)}_{n} \delta\left(p^2\right).
\end{eqnarray}

During deriving the chiral kinetic equations in seven-dimensional phase space or calculating some physical quantities such as charge currents and energy-momentum tensor, we always encounter some integrations over the Wigner function with a specific integral weight which is a function of $p_n$.  Therefore, we will integrate the Wigner function   over $p_n$ with a general weight function $g(p_n)$:
\begin{eqnarray}
\label{J-n-int-g}
\int dp_n g(p_n) \mathscr{J}^{(0)}_{n}&=&\frac{1}{2}\sum_\lambda g(\lambda |\bar p|) \lambda \mathcal{J}^{(0)}_{n}(\lambda ),\\
\label{J-bar-int-g}
\int dp_n g(p_n) \bar{\mathscr{J}}^{(0)\mu}&=&\frac{1}{2}\sum_\lambda g(\lambda |\bar p|) \hat{ \bar p}^\mu   \mathcal{J}^{(0)}_{n}(\lambda),
\end{eqnarray}
where the index $\lambda$ denotes positive ($\lambda=+1$) or negative ($\lambda=-1$) energy contribution and  $\mathcal{J}^{(0)}_{sn}(\lambda)$  and  $\hat{\bar p}^\mu $ are defined as
\begin{eqnarray}
 \mathcal{J}^{(0)}_{n}(\lambda ) \equiv  \left.\mathcal{J}^{(0)}_{n}\right|_{p_n=\lambda |\bar p|},
 \ \ \   \hat{ \bar p}^\mu \equiv \frac{\bar p^\mu}{|\bar p|}.
\end{eqnarray}
With the general expression, we can have specific results
\begin{eqnarray}
\label{Jn-0-mu-7d}
\int dp_n  \mathscr{J}^{(0)\mu}&=&\frac{1}{2}\sum_\lambda \sqrt{\omega}^{(0)} \dot{x}^{(0)\mu }  \mathcal{J}^{(0)}_{n},\\
\label{Jn-0-pn-7d}
\int dp_n p_n \mathscr{J}^{(0)}_{n}&=&\frac{1}{2}\sum_\lambda \sqrt{\omega}^{(0)}\varepsilon_n^{(0)} \mathcal{J}^{(0)}_{n}.
\end{eqnarray}
In the expressions above, we have defined some useful variables or  functions
\begin{eqnarray}
\sqrt{\omega}^{(0)} =1,\ \ \ \dot{x}^{(0)\mu} = \lambda n^\mu + \hat{\bar p}^\mu, \ \ \ \  \varepsilon^{(0)}_n =  |\bar p|,\ \ \ \
\end{eqnarray}
where $\sqrt{\omega}^{(0)}$ represents the factor associated with the  invariant phase space \cite{Son:2012zy} which is unity at zeroth order, $\dot{x}^{(0)\mu}$  the four-velocity for particles at
zeroth order which is just four-velocity for free chiral fermions, and $\varepsilon^{(0)}_n $  the energy for particles at zeroth order which is just onshell energy for free chiral fermions.
For brevity, we have suppressed above and will suppress below  all the arguments associated with $\lambda$ in $ \mathcal{J}^{(0)}_{n}$ or $\dot{x}^{(0)\mu}$.

The chiral kinetic equation at zeroth order could be obtained  directly by  integrating   the zeroth-order  Wigner equations (\ref{Js-ev-0}) over the momentum $p_n$.
The integral interval from $0$ to $+\infty$ gives the positive-energy contribution which is associated with the particle's distribution while the integral interval from $-\infty$ to $0$
gives the negative-energy contribution which is associated with the antiparticle's distribution.
However, it is valuable to integrate Eq.(\ref{Js-ev-0}) together with the arbitrary weight function $g(p_n)$, i.e.
\begin{eqnarray}
\label{Js-ev-7d-a}
\int dp_n g(p_n)\nabla_\mu\mathscr{J}^{(0)\mu} &=&0.
\end{eqnarray}
The terms including the derivative with respect to $p_n$ in the operator (\ref{nabla-bar}) can be integrated with the integral by parts while the derivatives with respect to $\bar p_\mu$ can be
directly pulled out of the integral
\begin{eqnarray}
\label{Js-ev-7d-b}
E_\mu \int dp_n g'(p_n)\mathscr{J}^{(0)\mu} +\nabla_\mu \int dp_n g(p_n) \mathscr{J}^{(0)\mu}   &=&0.
\end{eqnarray}
It should be pointed out that we have used the same symbol $\nabla_\mu$ to represent the corresponding operator where the derivative with respect to $p_n$ has been removed after we integrate over $p_n$, i.e., in the expression above
\begin{eqnarray}
\label{nabla-full-7d}
{\nabla}^\mu &=&\partial^x_\mu  +  E^\mu \bar\partial^p_\mu -\bar \epsilon^{\mu\rho\nu }B_\rho \bar \partial^p_\nu.
\end{eqnarray}
We expect the readers could easily distinguish them from the context.
It is remarkable that  the final result after integration is given by
\begin{eqnarray}
\label{CKE-0-7d-g}
\sum_{\lambda}g(\lambda |\bar p|)\nabla_\mu \left[\sqrt{\omega}^{(0)} \dot{x}^{(0)\mu }  \mathcal{J}^{(0)}_{n} \right] &=&0.
\end{eqnarray}
where  the terms including the derivative of $g(\lambda |\bar p|)$ have been cancelled. Since this equation holds for the arbitrary function $g(\lambda |\bar p|)$, it is only possible that
 the equation
\begin{eqnarray}
\label{CKE-0-7d}
\dot{x}^{(0)\mu } \nabla_\mu \mathcal{J}^{(0)}_{n}  &=&0,
\end{eqnarray}
must be satisfied for  positive and negative energy contributions separately. According to the correspondence between the positive/negative energy solution
and particle/antiparticle or direct calculation from the Wigner function definition  (\ref{wigner}) specific to a free particle, the CKE of the particles with momentum $\bar p^\mu $ and helicity $s$ can be obtained by setting $\lambda=+1$, while the CKE of the antiparticles with momentum $\bar p^\mu $ and helicity $s$ can be obtained by setting $\lambda=-1$ along with the replacement $\bar p^\mu \rightarrow -\bar p^\mu$ and $ s \rightarrow -s$. Such replacement
can be made for the Eqs.(\ref{J-n-int-g}), (\ref{J-bar-int-g}), (\ref{Jn-0-mu-7d}), and (\ref{Jn-0-pn-7d}).

\subsection{First-order CKT}

The Wigner tensor equations (\ref{Js-cs-1b}) with $ k\ge 1$ can be decomposed into   two groups of equations according to time-like or space-like components
\begin{eqnarray}
\label{Js-cs-1b-n}
\bar p^\mu \mathscr{J}_{n}^{(k)} - p_n \bar{\mathscr{J}}^{(k)\mu}
&=& -\frac{s}{2}\bar\epsilon^{\mu\rho\sigma}\bar\nabla_\rho \mathscr{J}_{\sigma}^{(k-1)},\\
\label{Js-cs-1b-bar}
\bar p^\mu \bar{\mathscr{J}}^{(k)\nu} - \bar p^\nu \bar{\mathscr{J}}^{(k)\mu}
&=& \frac{s}{2}\bar\epsilon^{\mu\nu\rho}\left(\nabla_n \bar{\mathscr{J}}_{\rho}^{(k-1)}
-\bar\nabla_\rho \mathscr{J}_{n}^{(k-1)}\right).
\end{eqnarray}
Setting $k=1$, the Eq.(\ref{Js-cs-1b-n}) relates $\bar{\mathscr{J}}^{(1)\mu}$ to $\mathscr{J}_{n}^{(1)}$ via
\begin{eqnarray}
\label{barJs-Jsn-1}
\bar{ \mathscr{J}}^{(1)\mu}
 &=&\frac{1}{p_n}\bar  p^\mu \mathscr{J}_{n}^{(1)} + \frac{s}{2p_n}\bar\epsilon^{\mu\rho\sigma}\nabla_\rho \mathscr{J}_{\sigma}^{(0)},
\end{eqnarray}
which implies that only the time-like component $\mathscr{J}_{n}^{(1)}$ is independent just similar to the case at zeroth order.
It is straightforward to verify that the equation (\ref{Js-cs-1b-bar}) is fulfilled automatically once  the Eqs.(\ref{barJs-Jsn-1}), (\ref{Js-cs-0a}), and (\ref{Js-ev-0}) hold \cite{Gao:2018wmr}.
Substituting the expression (\ref{barJs-Jsn-1}) into Eq.(\ref{Js-cs-1a}) with $k=1$ gives rise to
\begin{eqnarray}
 p^2 \frac{\mathscr{J}_{n}^{(1)} }{p_n }  &=&-\frac{s}{2p_n}\bar\epsilon^{\mu\rho\sigma}p_\mu \nabla_\rho \mathscr{J}_{\sigma}^{(0)}.
\end{eqnarray}
The non-vanishing contribution on the right-handed side implies the modification to the on-shell conditions for  free chiral particles
and the general solution must take  the form
\begin{eqnarray}
\label{Js-n-1}
\mathscr{J}^{(1)}_{n}&=& p_n  \tilde{\mathcal{J}}^{(1)}_{n} \delta\left(p^2\right)
-\frac{s}{2p^2}\bar\epsilon^{\mu\rho\sigma}p_\mu \nabla_\rho \mathscr{J}_{\sigma}^{(0)}\nonumber\\
&=&p_n  \tilde{\mathcal{J}}^{(1)}_{n} \delta\left(p^2\right) + s(B\cdot p)
\mathcal{J}^{(0)}_{n}\frac{ \delta\left(p^2\right)}{p^2},
\end{eqnarray}
where we have used the symbol  $\tilde{\mathcal{J}}^{(1)}_{n}$ because we will reserve the symbol $\mathcal{J}^{(1)}_{n}$ for further redefinition.
Plugging it into Eq.(\ref{barJs-Jsn-1}) leads to
\begin{eqnarray}
\label{Js-mu-1}
\bar{ \mathscr{J}}^{(1)\mu}
&=&\bar p^\mu \left[\tilde{\mathcal{J}}^{(1)}_{n} \delta\left(p^2\right)
+\frac{s B\cdot p}{ p_n }\mathcal{J}^{(0)}_{n} \frac{ \delta\left(p^2\right)}{p^2} \right]
 + \frac{s}{2p_n}\bar\epsilon^{\mu\rho\sigma}\nabla_\rho\left[ p_{\sigma}\mathcal{J}^{(0)}_{n} \delta\left(p^2\right) \right].
\end{eqnarray}
Following the similar approach used in zeroth order, we integrate $\mathscr{J}^{(1)}_{n}$  over $p_n$ with arbitrary weight function $g(p_n)$
\begin{eqnarray}
\label{J-1-n-g}
\int dp_n g(p_n)\mathscr{J}^{(1)}_{n}
&=&\frac{1}{2}\sum_\lambda g\left[\lambda \tilde{\mathcal{J}}^{(1)}_{n}
+\frac{s(B\cdot \bar p)}{2|\bar p|^2}\lambda \mathcal{J}^{(0)\prime}_{n}
- \frac{s(B\cdot \bar p)}{2|\bar p|^3} \mathcal{J}^{(0)}_{n}\right]\nonumber\\
& & +\frac{1}{2}\sum_\lambda g'\frac{s(B\cdot \bar p)}{2|\bar p|^2}\lambda \mathcal{J}^{(0)}_{n}.
\end{eqnarray}
We emphasize again that  we have suppressed all the arguments associated with $\lambda$  in the functions $g$, $\tilde{\mathcal{J}}^{(1)}_{n}$,
$\mathcal{J}^{(0)}_{n}$, and $\mathcal{J}^{(0)}_{n}$ as well as $\mathcal{J}^{(0)\prime}_{n}$ and $g'$
The prime on the functions $\mathcal{J}^{(0)}_{n}$ and $g$ denotes derivative as usual
\begin{eqnarray}
\mathcal{J}^{(0)\prime}_{n}\equiv  \left.\frac{\partial\mathcal{J}^{(0)}_{n}}{\partial p_n}\right|_{p_n=\lambda |\bar p|},\ \ \
 g'\equiv \left.\frac{d g(p_n)}{d p_n}\right|_{p_n=\lambda |\bar p|}.
\end{eqnarray}
From Eqs.(\ref{J-1-n-g}) and (\ref{Js-mu-1}), we obtain
\begin{eqnarray}
\label{barJ-1-g-7d}
\int dp_n g(p_n) \bar{\mathscr{J}}^{(1)\mu}
&=&\frac{1}{2}\sum_\lambda\left\{ g\hat{\bar p}^\mu  \left[\tilde{\mathcal{J}}^{(1)}_{n}
+\frac{s(B\cdot \bar p)}{2|\bar p|^2} \mathcal{J}^{(0)\prime}_{n}
-\frac{s(B\cdot \bar p)}{2|\bar p|^3}\lambda \mathcal{J}^{(0)}_{n}\right]\right.\nonumber\\
& & +\left( g' - \frac{\lambda g}{|\bar p|}\right)\hat{\bar p}^\mu
 \frac{s(B\cdot \bar p)}{2|\bar p|^2} \mathcal{J}^{(0)}_{n}
\nonumber\\
& &+ s\left(\frac{\lambda g'}{2|\bar p|^2}-\frac{g}{2|\bar p|^3 } \right)
\bar\epsilon^{\mu\rho\sigma}E_\rho {\bar p}_{\sigma} \mathcal{J}^{(0)}_{n}\nonumber\\
& &\left. + s\sum_\lambda \bar\epsilon^{\mu\rho\sigma}\bar \nabla_\rho
\left(\frac{\lambda g}{2|\bar p|^2}\bar p_{\sigma}\mathcal{J}^{(0)}_{n}  \right)\right\}.
\end{eqnarray}
Additional benefit by introducing an arbitrary weight function $g(p_n)$ is that we can easily recognize that some functions always  appear as a whole. As we all know, after integration over $p_n$,
there is no relation between $ \mathcal{J}^{(0)\prime}_{n}$ and $ \mathcal{J}^{(0)}_{n}$ any more. However, from the expression (\ref{J-1-n-g}), we note that $ \mathcal{J}^{(0)\prime}_{n}$ and
$ \tilde{\mathcal{J}}^{(1)}_{n}$ always appear together as a whole, we can redefine a new distribution function by
\begin{eqnarray}
\mathcal{J}^{(1)}_{n} \equiv  \tilde{\mathcal{J}}^{(1)}_{n}
+\frac{s(B\cdot \bar p)}{2|\bar p|^2} \mathcal{J}^{(0)\prime}_{n},
\end{eqnarray}
This redefinition has a very clear physical meaning: the second term is the first-order correction for $ \mathcal{J}^{(0)}_{n}$ due to  the energy shift ${s(B\cdot \bar p)}/{2|\bar p|^2}$ generated from the interaction between magnetic field and spin.
With this new distribution function and specific cases of $g(p_n)=1$ and $g(p_n)=p_n$,  we can have
\begin{eqnarray}
\int dp_n  {\mathscr{J}}_{n}^{(1)}
&=&\frac{1}{2}\sum_\lambda\lambda\left(\sqrt{\omega}^{(0)}\mathcal{J}^{(1)}_{n}
+ \sqrt{\omega}^{(1)} \mathcal{J}^{(0)}_{n} \right),\\
\int dp_n p_n {\mathscr{J}}^{(1)}_{n} &=&
\frac{1}{2}\sum_\lambda\left(\sqrt{\omega}^{(0)}\varepsilon^{(0)} \mathcal{J}^{(1)}_{n}
+ \sqrt{\omega}^{(0)} \varepsilon^{(1)}\tilde{\mathcal{J}}^{(0)}_{n}
 +\sqrt{\omega}^{(1)}  \varepsilon^{(0)}\mathcal{J}^{(0)}_{n} \right)
\end{eqnarray}
where
\begin{eqnarray}
\label{omega-1}
\sqrt{\omega}^{(1)} &=& -\frac{s\lambda(B\cdot \bar p)}{2|\bar p|^3},\ \ \ \
\varepsilon^{(1)} = \frac{s\lambda(B\cdot \bar p)}{2|\bar p|^2}
\end{eqnarray}
denotes the first-order modification to the invariant phase space and the particle's energy \cite{Son:2012zy}, respectively.  The chiral Wigner function after integration over $p_n$ can be written as
\begin{eqnarray}
\label{Jn-1-mu-7d}
\int dp_n \mathscr{J}^{(1)\mu}
&=&\frac{1}{2}\sum_\lambda\left[\sqrt{\omega}^{(0)}\dot{x}^{(0)\mu} \mathcal{J}^{(1)}_{n}
+ \sqrt{\omega}^{(1)} \dot{x}^{(0)\mu}  \mathcal{J}^{(0)}_{n}
+\sqrt{\omega}^{(0)}\dot{x}^{(1)\mu} \mathcal{J}^{(0)}_{n}\right]
\end{eqnarray}
where first-order modification to the four-velocity $\dot{x}^{(0)\mu}$ is given by
\begin{eqnarray}
\label{dot-x-1}
\dot{x}^{(1)\mu} &=& - \frac{s\lambda(B\cdot \bar p)}{2|\bar p|^3}\hat{\bar p}^\mu
- \frac{s}{2|\bar p|^3 }\bar\epsilon^{\mu\rho\sigma}E_\rho {\bar p}_{\sigma}
 + \frac{s \lambda }{ 2 }\bar\epsilon^{\mu\rho\sigma}\bar \nabla_\rho
\frac{\bar p_{\sigma}}{|\bar p|^2}.
\end{eqnarray}
It needs to  be emphasized that we should regard $\dot{x}^{(1)\mu}$ as a operator instead of a ordinary function. The operator $\bar \nabla_\rho$
in the last term of the Eq.(\ref{dot-x-1}) acts on not only ${\bar p_{\sigma}}/{|\bar p|^2}$ but also the distribution function $ \mathcal{J}^{(0)}_{n}$.
The first-order CKE in seven-dimensional phase space  can be derived from Eq.(\ref{Js-ev-1}) with $k=1$ by following the similar procedure in zeroth order from Eq.(\ref{Js-ev-7d-a}) to Eq.(\ref{CKE-0-7d-g}).
It turns out that all  the terms including the derivative of $g$ have been still cancelled  at first order just like the case at zeroth-order and
only an overall factor $g$ remains. It follows that the following equation must hold for both positive and negative energy parts separately:
\begin{eqnarray}
0&=& \nabla_\mu \left[\sqrt{\omega}^{(0)}\dot{x}^{(0)\mu} \mathcal{J}^{(1)}_{n}
+ \sqrt{\omega}^{(1)} \dot{x}^{(0)\mu}  \mathcal{J}^{(0)}_{n}
+\sqrt{\omega}^{(0)}\dot{x}^{(1)\mu} \mathcal{J}^{(0)}_{n}\right].
\end{eqnarray}
We can move the operator $\nabla_\mu$ adjacent to the distribution function
\begin{eqnarray}
0 &=& \sqrt{\omega}^{(0)}\dot{x}^{(0)\mu } \nabla_\mu \mathcal{J}^{(1)}_{n}
+  \sqrt{\omega}^{(1)} \dot{x}^{(0)\mu}  \nabla_\mu \mathcal{J}^{(0)}_{n}
+ \sqrt{\omega}^{(0)}\dot{x}^{(1)\mu } \nabla_\mu \mathcal{J}^{(0)}_{n} \nonumber\\
& & + \left[\nabla_\mu , \sqrt{\omega}^{(1)}\dot{x}^{(0)\mu} \right] \mathcal{J}^{(0)}_{n}
+ \left[\nabla_\mu , \sqrt{\omega}^{(0)}\dot{x}^{(1)\mu} \right] \mathcal{J}^{(0)}_{n}.
\end{eqnarray}
By utilizing  the zeroth-order CKE (\ref{CKE-0-7d}) for the second term on the right-handed side of the equation above, we have
\begin{eqnarray}
\label{CKE-1-7d}
0 &=&\dot{x}^{(0)\mu } \nabla_\mu \mathcal{J}^{(1)}_{n}
+\dot{x}^{(1)\mu } \nabla_\mu \mathcal{J}^{(0)}_{n}
 +\dot{x}^{(0)\mu}  \left[\nabla_\mu , \sqrt{\omega}^{(1)}\right] \mathcal{J}^{(0)}_{n}
+ \left[\nabla_\mu , \dot{x}^{(1)\mu} \right] \mathcal{J}^{(0)}_{n}.
\end{eqnarray}
Substituting the specific results for  $\sqrt{\omega}^{(1)}$ in Eq.(\ref{omega-1}) and  $\dot{x}^{(1)\mu } $ in Eq. (\ref{dot-x-1}) into the equation above and
dealing with some vector analysis, we obtain
\begin{eqnarray}
0&=& \dot{x}^{(0)\mu}\nabla_\mu  \mathcal{J}^{(1)}_{n}
-  \frac{s B\cdot \bar p}{2|\bar p|^3} \hat{ \bar p}^\mu \bar \nabla_\mu {\mathcal{J}}^{(0)}_{n}
- \frac{\lambda s}{2|\bar p|^3}\bar\epsilon^{\mu\rho\sigma} E_\rho  {\bar p}_\sigma
\bar \nabla_\mu {\mathcal{J}}^{(0)}_{n}.
\end{eqnarray}
Remember that the operator $\nabla_\mu$  includes  only  derivatives on $\bar p^\mu$ after integration over $p_n$ and the CKE for the particles with momentum $\bar p^\mu $
and helicity $s$  is obtained by setting $\lambda=+1$, while the CKE for the antiparticles with momentum $\bar p^\mu $ and helicity $s$  by setting $\lambda=-1$ along with the replacement $\bar p^\mu \rightarrow -\bar p^\mu$ and $ s \rightarrow -s$.

\subsection{Second-order CKT}

Now we come to our major task of  deriving second-order CKT in seven-dimensional phase space.  From  the second-order Wigner equation (\ref{Js-cs-1b-n}) by setting $k=2$, we can express $\bar{ \mathscr{J}}^{(2)\mu}$ in terms of $\mathscr{J}_{n}^{(2)}$ via
\begin{eqnarray}
\label{barJs-Jsn-2}
\bar{ \mathscr{J}}^{(2)\mu}
 &=&\frac{1}{p_n}\bar  p^\mu \mathscr{J}_{n}^{(2)} + \frac{s}{2p_n}\bar\epsilon^{\mu\rho\sigma}\nabla_\rho \mathscr{J}_{\sigma}^{(1)}.
\end{eqnarray}
Similar to the case at zeroth and first order, we can verify that the equation (\ref{Js-cs-1b-bar}) holds automatically  given the   Eqs.(\ref{barJs-Jsn-2}) and (\ref{barJs-Jsn-1}) together with  the specific case $k=1$ for (\ref{Js-cs-1a}) and (\ref{Js-ev-1}). Actually such remarkable procedure can be generalized
to any higher order and has been verified  systematically  in Ref.\cite{Gao:2018wmr}.   Substituting the expression (\ref{barJs-Jsn-1}) into Eq.(\ref{Js-cs-1a}) with $k=2$ gives rise to
\begin{eqnarray}
 p^2 \frac{\mathscr{J}_{n}^{(2)} }{p_n }  &=&-\frac{s}{2p_n}\bar\epsilon^{\mu\rho\sigma}p_\mu \nabla_\rho \mathscr{J}_{\sigma}^{(1)},
 \end{eqnarray}
which implies the correction to on-shell condition of the second-order Wigner function $\mathscr{J}_{n}^{(2)} $.
 Inserting the expressions (\ref{Js-n-1}) and (\ref{Js-mu-1}) into the right-hand side of the equation above, the general expression for $\mathscr{J}_{sn}^{(2)}$ must take the form
\begin{eqnarray}
\label{Jn-2-0}
\mathscr{J}^{(2)}_{n}&=& p_n \tilde{ \mathcal{J}}^{(2)}_{n} \delta\left(p^2\right)
-\frac{s}{2p^2}\bar\epsilon^{\mu\rho\sigma}p_\mu \nabla_\rho \mathscr{J}_{\sigma}^{(1)}
\nonumber\\
&=&  p_n \tilde{\mathcal{J}}^{(2)}_{n} \delta\left(p^2\right)
+ \frac{s B\cdot p}{p^2} \left[ \tilde{ \mathcal{J}}^{(1)}_{n} \delta\left(p^2\right) + \frac{s B\cdot p}{p_n}
\mathcal{J}^{(0)}_{n}\frac{ \delta\left(p^2\right)}{p^2}\right]\nonumber\\
& &-\frac{s}{2p^2}\bar\epsilon^{\mu\rho\sigma}p_\mu \nabla_\rho \left\{
\frac{s}{2p_n}\bar\epsilon_{\sigma\alpha\beta}\nabla^\alpha\left[ p^{\beta}\mathcal{J}^{(0)}_{n} \delta\left(p^2\right) \right] \right\}
\nonumber\\
&=&  p_n  \tilde{\mathcal{J}}^{(2)}_{n} \delta\left(p^2\right)
+ \frac{s B\cdot p}{p^2} \left[ \tilde{ \mathcal{J}}^{(1)}_{n} \delta\left(p^2\right) + \frac{s B\cdot p}{p_n}
\mathcal{J}^{(0)}_{n}\frac{ \delta\left(p^2\right)}{p^2}\right]\nonumber\\
& &+\frac{s}{2p^2}\left(\bar p_\mu \bar\nabla_\rho - \bar p_\rho \bar\nabla_\mu\right)
\left\{ \frac{s}{2p_n}\nabla^\mu\left[ p^{\rho}\mathcal{J}^{(0)}_{n} \delta\left(p^2\right) \right] \right\},
\end{eqnarray}
where we have used the identity
\begin{eqnarray}
\bar\epsilon^{\mu\nu\alpha}{\bar\epsilon}_{\alpha\rho\sigma}
&=& -{\Delta^{\mu}}_\rho {\Delta^{\nu}}_\sigma + {\Delta^{\nu}}_\rho {\Delta^{\mu}}_\sigma.
\end{eqnarray}
It is much more complicated and difficult to carry out the integral over $p_n$ for $\mathscr{J}^{(2)}_{n}$ together with a weight function $g(p_n)$. The more details can be found in Appendix \ref{identities}. Here we only present the final result:
\begin{eqnarray}
\label{Js-2-g-7d}
 & & \int dp_n g(p_n){\mathscr{J}}^{(2)}_{n} \nonumber\\
&=&  \frac{1}{2}\sum_\lambda\left\{ \lambda g\left(\mathcal{J}^{(2)}_{n}
+ \sqrt{\omega}^{(1)} \mathcal{J}^{(1)}_{n}
+\sqrt{\omega}^{(2)}\mathcal{J}^{(0)}_{sn} \right)
+\frac{1}{2}\lambda g'\frac{s B\cdot \bar p}{|\bar p|^2}
  \mathcal{J}^{(1)}_{n}\right.\nonumber\\
& &\hspace{1cm}+\frac{1}{{24}} \left(\frac{ g'''}{|\bar p|^3}
-\frac{{3}\lambda g''}{|\bar p|^4}
+{ \frac{{3}g' }{|\bar p|^5}} \right) \left[(B\cdot \bar p)^2 -B^2 \bar p^2\right]
  \mathcal{J}^{(0)}_{n} \nonumber\\
& &\hspace{1cm}+\frac{1}{{24}}
\left(\frac{g'''}{|\bar p|^3}  - \frac{3\lambda g''}{|\bar p|^4}
+ \frac{3 g'}{|\bar p|^5}  \right)\left[(E\cdot \bar p)^2 -\bar p^2 E^2\right]\mathcal{J}^{(0)}_{n} \nonumber\\
& &\hspace{1cm}+\frac{1}{12}\left( \frac{\lambda g'''}{|\bar p|^2}  - \frac{{3} g''}{|\bar p|^3}
+  \frac{{6}\lambda g'}{|\bar p|^4} \right) \bar \epsilon^{\mu\rho\sigma} E_\mu B_\rho \bar p_\sigma\mathcal{J}^{(0)}_{n}\nonumber\\
& &\hspace{1cm}+\frac{1}{8}\left(\frac{g''}{|\bar p|^3} -\frac{2\lambda g'}{|\bar p|^4}  \right)
\left[ (E\cdot\bar p) \bar p^{\mu} -\bar p^2 E^{\mu} + \lambda \bar \epsilon^{\mu\rho\sigma} B_\rho \bar p_\sigma \right]
\bar \nabla_\mu\mathcal{J}^{(0)}_{n} \nonumber\\
& &\left.\hspace{1cm}+\frac{1}{8} \frac{ g' }{|\bar p|^3} \left(\bar p_\mu \bar p_{\nu} -\bar p^2 \Delta_{\mu\nu}\right)
\bar\nabla^\mu \bar \nabla^\nu \mathcal{J}^{(0)}_{n}\right\},
\end{eqnarray}
where $\sqrt{\omega}^{(2)}$ can be regarded as the second-order correction to the invariant phase space
\begin{eqnarray}
\sqrt{\omega}^{(2)} & = &
- \frac{1}{2|\bar p|^5}\bar\epsilon^{\rho\mu\nu}\bar p_\rho E_\mu B_\nu
 -\frac{\lambda}{4} \bar\nabla^\mu  \bar \nabla^\nu  \frac{\bar p_\mu \bar p_{\nu} -\bar p^2 \Delta_{\mu\nu}}{|\bar p|^4}.
\end{eqnarray}
Just like $\dot{x}^{(1)\mu}$,  $\sqrt{\omega}^{(2)}$ is also identified as an operator because the operators  $\bar\nabla$  in the second term on the right-hand side of the equal sign above actually  act on all the terms  following it  including the distribution
function $\mathcal{J}^{(0)}_{n}$. Besides, we have defined a new distribution function $\mathcal{J}^{(2)}_{n}$ via
\begin{eqnarray}
\mathcal{J}^{(2)}_{n} &=& \tilde{\mathcal{J}}^{(2)}_{n}
+\frac{s B\cdot \bar p}{2|\bar p|^2} \tilde{\mathcal{J}}^{(1)\prime}_{n}
+\frac{1}{2}\left(\frac{s B\cdot \bar p}{2|\bar p|^2}\right)^2 \mathcal{J}^{(0)\prime\prime}_{n}\\
& &+\frac{1}{24}(B\cdot \bar p)^2\left(\frac{\lambda}{|\bar p|^3}\mathcal{J}^{(0)\prime\prime\prime}_{n}
- \frac{9\lambda }{|\bar p|^5}\mathcal{J}^{(0)\prime}_{n}\right)\nonumber\\
& &+\frac{1}{24}B^2\left(\frac{\lambda}{|\bar p|}\mathcal{J}^{(0)\prime\prime\prime}_{n}
-\frac{ 3 }{|\bar p|^2}\mathcal{J}^{(0)\prime\prime}_{n}
- \frac{3\lambda }{|\bar p|^3}\mathcal{J}^{(0)\prime}_{n}\right)\nonumber\\
& &+\frac{1}{24}\left[(E\cdot\bar p)^2 -E^2 \bar p^2 \right]
\left( \frac{\lambda}{|\bar p|^3 }\mathcal{J}^{(0)\prime\prime\prime}_{n}
-\frac{3 }{|\bar p|^4 }\mathcal{J}^{(0)\prime\prime}_{n}
+\frac{3 \lambda}{|\bar p|^5 }\mathcal{J}^{(0)\prime}_{n}\right)\nonumber\\
& &+\frac{1}{12}\bar\epsilon^{\rho\mu\nu}\bar p_\rho E_\mu B_\nu
\left( \frac{1}{|\bar p|^2 }\mathcal{J}^{(0)\prime\prime\prime}_{n}
-\frac{3\lambda }{|\bar p|^3 }\mathcal{J}^{(0)\prime\prime}_{n}
+\frac{3 }{|\bar p|^4 }\mathcal{J}^{(0)\prime}_{n}\right)\nonumber\\
& &-\frac{1}{8}
\left[ (E\cdot \bar p) \bar p^{\mu} -\bar p^2 E^\mu +\lambda|\bar p| \bar \epsilon^{\mu\nu\lambda}B_{\nu}  \bar p_\lambda\right]
\left(\frac{\lambda}{|\bar p|^3}\bar \nabla_\mu\mathcal{J}^{(0)\prime\prime}_{n}
-\frac{2 }{|\bar p|^4}\bar \nabla_\mu\mathcal{J}^{(0)\prime}_{n} \right)\nonumber\\
& &+\frac{\lambda}{8|\bar p|^3} \left(\bar p_\mu \bar p_{\nu} -\bar p^2 \Delta_{\mu\nu}\right)
\bar\nabla^\mu \bar \nabla^\nu \mathcal{J}^{(0)\prime}_{n},
\end{eqnarray}
 where the second term on the right-handed side of the first line above can be identified as the first-order correction for
 $ \mathcal{J}^{(1)}_{n}$ while the third term as the second-order correction for $ \mathcal{J}^{(1)}_{n}$ due to  the energy shift ${s(B\cdot \bar p)}/{2|\bar p|^2}$. We can identify the terms associated with $\mathcal{J}^{(0)\prime}_{n}$ as the second-order correction for
 the energy shift. However, all  other terms associated with $\mathcal{J}^{(0)\prime\prime\prime}_{n}$, $\mathcal{J}^{(0)\prime\prime}_{n}$,
 or operator $\bar\nabla^\mu$ have no explicit intuitive physical meaning except that the term of last line originate from the uncertainty
 principle. These terms have not been obtained in previous work.

For the specific cases $g(p_n)=1$ and $g(p_n)=p_n$, we have
\begin{eqnarray}
\label{Js-2-1-7d}
\int dp_n  {\mathscr{J}}_{n}
&=&\frac{1}{2}\lambda\left[\sqrt{\omega}^{(0)} \tilde{\mathcal{J}}^{(2)}_{n}
+ \sqrt{\omega}^{(1)} \tilde{\mathcal{J}}^{(1)}_{n}
+\sqrt{\omega}^{(2)}\mathcal{J}^{(0)}_{n} \right],
\end{eqnarray}
and
\begin{eqnarray}
\label{Js-2-pn-7d}
\int dp_n p_n {\mathscr{J}}^{(2)}_{n} &=&
\frac{1}{2}\left[\sqrt{\omega}^{(0)}\varepsilon^{(0)} \tilde{\mathcal{J}}^{(2)}_{n}
+ \sqrt{\omega}^{(1)} \varepsilon^{(0)}\tilde{\mathcal{J}}^{(1)}_{n}
+ \sqrt{\omega}^{(0)} \varepsilon^{(1)}\tilde{\mathcal{J}}^{(1)}_{n}\right.\nonumber\\
& &\left.\hspace{1cm} +\sqrt{\omega}^{(1)}  \varepsilon^{(1)} \mathcal{J}^{(0)}_{n}
+\sqrt{\omega}^{(0)}  \varepsilon^{(2)} \mathcal{J}^{(0)}_{n}
+\sqrt{\omega}^{(2)}  \varepsilon^{(0)}\mathcal{J}^{(0)}_{n} \right],
\end{eqnarray}
where $ \varepsilon^{(2)}$ can be interpreted as  second-order correction to the energy for  free chiral fermions
\begin{eqnarray}
\label{epsilon}
\varepsilon^{(2)} &=& \frac{1}{{8|\bar p|^5}} \left[(B\cdot \bar p)^2 + B^2 \bar p^2 + (E\cdot \bar p)^2 -\bar p^2 E^2 \right]
+\frac{\lambda}{4|\bar p|^4} \bar \epsilon^{\mu\rho\sigma} E_\mu B_\rho \bar p_\sigma \nonumber\\
& & -\frac{\lambda}{8} \bar \nabla_\mu
\frac{(E\cdot p) \bar p^{\mu}  -\bar p^2 E^{\mu} -\lambda |\bar p| \bar\epsilon^{\mu\rho\sigma}B_\rho \bar p_\sigma}{|\bar p|^4}
+\frac{1 }{8} \bar\nabla^\mu \bar \nabla^\nu \frac{\bar p_\mu \bar p_{\nu} -\bar p^2 \Delta_{\mu\nu}}{ |\bar p|^3}.
\end{eqnarray}
Just like $\sqrt{\omega}^{(2)}$ and $\dot{x}^{(1)\mu}$, the terms  with $\bar\nabla$ denote operators which act on all the terms following it.
Now let  us calculate the integral of the chiral Wigner function $\mathscr{J}^{(2)\mu}$
\begin{eqnarray}
\label{Jn-2-mu-7d}
\int dp_n  \mathscr{J}^{(2)\mu}
&=&\int dp_n \left[\frac{1}{p_n}  p^\mu \mathscr{J}_{n}^{(2)}
 + \frac{s}{2p_n}\bar\epsilon^{\mu\rho\sigma}\nabla_\rho \bar{\mathscr{J}}_{\sigma}^{(1)}\right]\nonumber\\
&=&n^\mu \int dp_n \mathscr{J}_{n}^{(2)} + \bar p^\mu \int dp_n \frac{1}{p_n} \mathscr{J}_{n}^{(2)}
\nonumber\\
& & - \frac{s}{2}\bar\epsilon^{\mu\rho\sigma}E_\rho \int dp_n \frac{1}{p_n^2}\bar{\mathscr{J}}_{\sigma}^{(1)}
 + \frac{s}{2}\bar\epsilon^{\mu\rho\sigma}\nabla_\rho \int dp_n \frac{1}{p_n}\bar{\mathscr{J}}_{\sigma}^{(1)}.
\end{eqnarray}
The first term after the last equal sign can be obtained by Eq.(\ref{Js-2-1-7d}), the second term  by setting $g(p_n)=1/p_n$ in Eq.(\ref{Js-2-g-7d}), and
the third and fourth terms by setting  $g(p_n)=1/p_n^2$ and $g(p_n)=1/p_n$ in Eq.(\ref{barJ-1-g-7d}), respectively.
The final result can be organized into
\begin{eqnarray}
\int dp_n  \mathscr{J}^{(2)\mu} &=&
\frac{1}{2}\left[\sqrt{\omega}^{(0)}\dot{x}^{(0)\mu} {\mathcal{J}}^{(2)}_{n}
+ \sqrt{\omega}^{(1)}\dot{x}^{(0)\mu}{\mathcal{J}}^{(1)}_{n}
+ \sqrt{\omega}^{(0)} \dot{x}^{(1)\mu}{\mathcal{J}}^{(1)}_{n}\right.\nonumber\\
& &\left.\hspace{1cm} +\sqrt{\omega}^{(1)} \dot{x}^{(1)\mu} \mathcal{J}^{(0)}_{n}
+ \sqrt{\omega}^{(0)}  \dot{x}^{(2)\mu} \mathcal{J}^{(0)}_{n} +\sqrt{\omega}^{(2)} \dot{x}^{(0)\mu}\mathcal{J}^{(0)}_{n} \right],
\end{eqnarray}
where $\dot{{x}}^{(2)\mu}$ denotes second-order correction to the four-velocity $\dot{x}^{(0)\mu}$ and actually represents an operator

\begin{eqnarray}
\dot{{x}}^{(2)\mu}  & \equiv & \frac{3|\bar p|(E\cdot p) E^\mu -5 |\bar p|(B\cdot p) B^\mu
+  \lambda(B\cdot p)\bar\epsilon^{\mu \alpha\beta } E_\alpha \bar p_\beta }{8|\bar p|^6 }
\nonumber\\
& &-\frac{11(B\cdot \bar p)^2 +B^2 \bar p^2
+5(E\cdot \bar p)^2 -11\bar p^2 E^2 + 4\lambda |\bar p| \bar \epsilon^{\nu\alpha\beta} E_\nu B_\alpha \bar p_\beta}{{16}|\bar p|^6}\hat{\bar p}^\mu
\nonumber\\
& & -\bar\nabla_\alpha \bar \nabla_\beta  \frac{ \bar p^\alpha \bar p^{\beta}\bar p^\mu -\bar p^2\bar p^\mu \Delta^{\alpha\beta}
 +2\bar p^2 \bar p^\mu \Delta^{\alpha\beta} -2\bar p^2 \bar p^\alpha \Delta^{\mu\beta} }{16|\bar p|^5} \nonumber\\
& &+\lambda E_\alpha \bar \nabla_\beta
\frac{2\bar p^\alpha \bar p^{\beta}{\bar p}^\mu + \bar p^2{\bar p}^\mu \Delta^{\alpha\beta}
-\bar p^2 \bar p^\alpha \Delta^{\mu\beta} -2\bar p^2 \bar p^\beta \Delta^{\mu\alpha} }{8|p|^6}
 \nonumber\\
& &+  B_\alpha \bar \nabla_\beta\frac{\bar \epsilon^{\beta\alpha\nu} \bar p_\nu \bar p^\mu + 2 \bar\epsilon^{\mu\beta\nu}\bar p_\nu \bar p^\alpha }{8|\bar p|^5}
\end{eqnarray}
Following the similar procedure at   zeroth or first order, second-order CKE  is given by
\begin{eqnarray}
0&=& \nabla_\mu\left[\sqrt{\omega}^{(0)}\dot{x}^{(0)\mu}\mathcal{J}^{(2)}_{n}
+ \sqrt{\omega}^{(1)}\dot{x}^{(0)\mu}\mathcal{J}^{(1)}_{n}
+ \sqrt{\omega}^{(0)} \dot{x}^{(1)\mu}\mathcal{J}^{(1)}_{n}\right.\nonumber\\
& &\left.\hspace{1cm} +\sqrt{\omega}^{(1)} \dot{x}^{(1)\mu} \mathcal{J}^{(0)}_{n}
+ \sqrt{\omega}^{(0)}  \dot{x}^{(2)\mu} \mathcal{J}^{(0)}_{n} +\sqrt{\omega}^{(2)} \dot{x}^{(0)\mu}\mathcal{J}^{(0)}_{n} \right]
\end{eqnarray}
Moving the operator $ \nabla_\mu$ all through $\sqrt{\omega}$ and $\dot{x}^{\mu}$ to $\mathcal{J}_{n}$ gives rise to
\begin{eqnarray}
0&=&\sqrt{\omega}^{(0)}\dot{x}^{(0)\mu}\nabla_\mu \mathcal{J}^{(2)}_{n}
+ \sqrt{\omega}^{(1)}\dot{x}^{(0)\mu}\nabla_\mu\mathcal{J}^{(1)}_{n}
+ \sqrt{\omega}^{(0)} \dot{x}^{(1)\mu}\nabla_\mu\mathcal{J}^{(1)}_{n}\nonumber\\
& & +\sqrt{\omega}^{(1)} \dot{x}^{(1)\mu}\nabla_\mu \mathcal{J}^{(0)}_{n}
+ \sqrt{\omega}^{(0)}  \dot{x}^{(2)\mu}\nabla_\mu \mathcal{J}^{(0)}_{n} +\sqrt{\omega}^{(2)} \dot{x}^{(0)\mu}\nabla_\mu\mathcal{J}^{(0)}_{n} \nonumber\\
& &+\left[\nabla_\mu, \sqrt{\omega}^{(0)}\dot{x}^{(0)\mu}\right]\mathcal{J}^{(2)}_{n}
+ \left[\nabla_\mu,\sqrt{\omega}^{(1)}\dot{x}^{(0)\mu}\right] \mathcal{J}^{(1)}_{n}
+ \left[\nabla_\mu, \sqrt{\omega}^{(0)} \dot{x}^{(1)\mu}\right]\mathcal{J}^{(1)}_{n}\nonumber\\
& & +\left[\nabla_\mu,\sqrt{\omega}^{(1)} \dot{x}^{(1)\mu}\right] \mathcal{J}^{(0)}_{n}
+ \left[\nabla_\mu , \sqrt{\omega}^{(0)}  \dot{x}^{(2)\mu}\right] \mathcal{J}^{(0)}_{n}
+\left[\nabla_\mu,\sqrt{\omega}^{(2)} \dot{x}^{(0)\mu}\right]\mathcal{J}^{(0)}_{n}.
\end{eqnarray}
Using the zeroth-order and first-order  CKE (\ref{CKE-0-7d}) and (\ref{CKE-1-7d}), we can simplify the second-order CKE to
\begin{eqnarray}
\label{CKE-2-7d}
0&=&\dot{x}^{(0)\mu}\nabla_\mu \mathcal{J}^{(2)}_{n}
+  \dot{x}^{(1)\mu}\nabla_\mu\mathcal{J}^{(1)}_{n}
+   \dot{x}^{(2)\mu}\nabla_\mu \mathcal{J}^{(0)}_{n} \nonumber\\
& &+\left[\nabla_\mu, \dot{x}^{(0)\mu}\right] \mathcal{J}^{(2)}_{n}
+ \left[\nabla_\mu,\sqrt{\omega}^{(1)}\dot{x}^{(0)\mu}\right] \mathcal{J}^{(1)}_{n}
+ \left[\nabla_\mu, \dot{x}^{(1)\mu}\right]\mathcal{J}^{(1)}_{n}\nonumber\\
& & +\left[\nabla_\mu,\sqrt{\omega}^{(1)} \dot{x}^{(1)\mu}\right] \mathcal{J}^{(0)}_{n}
+ \left[\nabla_\mu , \dot{x}^{(2)\mu}\right] \mathcal{J}^{(0)}_{n}
+\left[\nabla_\mu,\sqrt{\omega}^{(2)} \dot{x}^{(0)\mu}\right]\mathcal{J}^{(0)}_{n}\nonumber\\
& & -\sqrt{\omega}^{(1)} \dot{x}^{(0)\mu}  \left[\nabla_\mu , \sqrt{\omega}^{(1)}\right] \mathcal{J}^{(0)}_{n}
-\sqrt{\omega}^{(1)} \left[\nabla_\mu , \dot{x}^{(1)\mu} \right] \mathcal{J}^{(0)}_{n}.
\end{eqnarray}
Substituting the specific expressions for $\dot{x}^{(1)\mu}$,  $\dot{x}^{(2)\mu}$, $\sqrt{\omega}^{(1)} $ and $\sqrt{\omega}^{(2)} $   into the formal result above, we obtain
the final second-order CKE
\begin{eqnarray}
0&=& \dot{x}^{(0)\mu}\nabla_\mu \mathcal{J}^{(2)}_{n}
-  \frac{s B\cdot \bar p}{2|\bar p|^3} \hat{ \bar p}^\mu \bar \nabla_\mu \mathcal{J}^{(1)}_{n}
- \frac{\lambda s}{2|\bar p|^3}\bar\epsilon^{\mu\rho\sigma} E_\rho  {\bar p}_\sigma
\bar \nabla_\mu \mathcal{J}^{(1)}_{n}\nonumber\\
& & +\frac{3(B\cdot \bar p)^2 -\bar p^2 B^2 - 5(E\cdot \bar p)^2 - \bar p^2 E^2
 - 4\lambda |\bar p| \bar\epsilon^{\nu\rho\sigma}{\bar p}_\nu E_\rho B_\sigma }{8|\bar p|^6}
  \lambda \hat{ \bar p}^\mu\bar\nabla_\mu  \mathcal{J}^{(0)}_{n}\nonumber\\
& &+ \frac{(B\cdot \bar p)}{4|\bar p|^6} \bar\epsilon^{\mu\rho\sigma} E_\rho {\bar p}_\sigma  \bar \nabla_\mu  \mathcal{J}^{(0)}_{n}
 - \frac{3(E\cdot \bar p)  }{4|\bar p|^5}\lambda E^\mu \bar\nabla_\mu \mathcal{J}^{(0)}_{n}\nonumber\\
& & -\frac{ \bar p_\rho \bar p_{\sigma} -\bar p^2 \Delta_{\rho\sigma}}{8|\bar p|^4}\lambda\hat{ \bar p}^\mu\bar\nabla_\mu
\bar\nabla^\rho \bar \nabla^\sigma \mathcal{J}^{(0)}_{n}
+\frac{\bar \epsilon^{\nu\rho\sigma} \bar p_\rho  B_\sigma }{4|\bar p|^4}
 \lambda\hat{ \bar p}^\mu\bar\nabla_\mu\bar \nabla_\nu \mathcal{J}^{(0)}_{n}\nonumber\\
& &+\frac{2(E\cdot\bar p)(\bar p^2 \Delta^{\mu\nu} - \bar p^{\mu} \bar p^{\nu} )
+\bar p^2 \left[\bar p^\mu E^{\nu} - (E\cdot \bar p) \Delta^{\mu\nu}\right]}{4|\bar p|^6}
\bar\nabla_\mu \bar \nabla_\nu  \mathcal{J}^{(0)}_{n}.
\end{eqnarray}

\subsection{The summed  CKT }

In previous three subsections, we provide a systematic method to derive the CKT in seven-dimensional phase space from the original Wigner equations in eight-dimensional phase space from zeroth to second order.
This procedure is very consistent and can be generalized to any higher order  systematically.  At the end of this section, we will recombine  these perturbative result  into a unified form.
To do that, we can define full distribution function in seven-dimensional phase space
\begin{eqnarray}
\label{Js-7d-expansion}
\mathcal{J}_n &=& \mathcal{J}^{(0)}_n +\hbar \mathcal{J}^{(1)}_n +\hbar^2  \mathcal{J}^{(2)}_n  +... ...
\end{eqnarray}
Then  in the light of the expressions (\ref{Jn-0-mu-7d}),  (\ref{Jn-1-mu-7d}), and  (\ref{Jn-2-mu-7d}),  we can write  the integration of  the full $\mathscr{J}^{\mu}$ as
\begin{eqnarray}
\int dp_n \mathscr{J}^{\mu}
&=&\frac{1}{2}\sqrt{\omega}\dot{x}^{\mu} \mathcal{J}_{n}.
\end{eqnarray}
The full CKE in seven-dimensional phase space will be given by
\begin{eqnarray}
0 &=& \nabla_\mu \left( \sqrt{\omega}\dot{x}^{\mu} \mathcal{J}_{n} \right),
\end{eqnarray}
or in another  form
\begin{eqnarray}
\label{CKE-full}
0 &=&\dot{x}^{\mu} \nabla_\mu  \mathcal{J}_{n} + \sqrt{\omega}^{-1}\left[\nabla_\mu , \sqrt{\omega}\dot{x}^{\mu} \right]  \mathcal{J}_{n}.
\end{eqnarray}
When we expand this equations order by order, we will reproduce the results given in Eqs.(\ref{CKE-0-7d}), (\ref{CKE-1-7d}) and (\ref{CKE-2-7d}).
Hence the  equation  (\ref{CKE-full}) can be identified as the full  equation of CKE in seven-dimensional phase space. Up to second order, we have
\begin{eqnarray}
\label{omega-full}
\sqrt{\omega} & = &1-\frac{s\lambda(B\cdot \bar p)}{2|\bar p|^3}
- \frac{1}{2|\bar p|^5}\bar\epsilon^{\rho\mu\nu}\bar p_\rho E_\mu B_\nu
 -\frac{\lambda}{4} \bar\nabla^\mu  \bar \nabla^\nu  \frac{\bar p_\mu \bar p_{\nu} -\bar p^2 \Delta_{\mu\nu}}{|\bar p|^4},\\
\label{epsilon-full}
\varepsilon &=&|\bar p|+\frac{s\lambda(B\cdot \bar p)}{2|\bar p|^2}+ \frac{1}{{8|\bar p|^5}} \left[(B\cdot \bar p)^2 + B^2 \bar p^2 + (E\cdot \bar p)^2 -\bar p^2 E^2 \right]
+\frac{\lambda}{4|\bar p|^4} \bar \epsilon^{\mu\rho\sigma} E_\mu B_\rho \bar p_\sigma \nonumber\\
& & -\frac{\lambda}{8} \bar \nabla_\mu
\frac{(E\cdot p) \bar p^{\mu}  -\bar p^2 E^{\mu} -\lambda |\bar p| \bar\epsilon^{\mu\rho\sigma}B_\rho \bar p_\sigma}{|\bar p|^4}
+\frac{1 }{8} \bar\nabla^\mu \bar \nabla^\nu \frac{\bar p_\mu \bar p_{\nu} -\bar p^2 \Delta_{\mu\nu}}{ |\bar p|^3},\\
\label{velocity-full}
\dot{{x}}^{\mu}  & = &  \lambda n^\mu + \hat{\bar p}^\mu
- \frac{s\lambda(B\cdot \bar p)}{2|\bar p|^3}\hat{\bar p}^\mu
- \frac{s}{2|\bar p|^3 }\bar\epsilon^{\mu\rho\sigma}E_\rho {\bar p}_{\sigma}
 + \frac{s \lambda }{ 2 }\bar\epsilon^{\mu\rho\sigma}\bar \nabla_\rho
\frac{\bar p_{\sigma}}{|\bar p|^2}\nonumber\\
& &+\frac{3|\bar p|(E\cdot p) E^\mu -5 |\bar p|(B\cdot p) B^\mu
+  \lambda(B\cdot p)\bar\epsilon^{\mu \alpha\beta } E_\alpha \bar p_\beta }{8|\bar p|^6 }
\nonumber\\
& &-\frac{11(B\cdot \bar p)^2 +B^2 \bar p^2
+5(E\cdot \bar p)^2 -11\bar p^2 E^2 + 4\lambda |\bar p| \bar \epsilon^{\nu\alpha\beta} E_\nu B_\alpha \bar p_\beta}{{16}|\bar p|^6}\hat{\bar p}^\mu
\nonumber\\
& & -\bar\nabla_\alpha \bar \nabla_\beta  \frac{ \bar p^\alpha \bar p^{\beta}\bar p^\mu -\bar p^2\bar p^\mu \Delta^{\alpha\beta}
 +2\bar p^2 \bar p^\mu \Delta^{\alpha\beta} -2\bar p^2 \bar p^\alpha \Delta^{\mu\beta} }{16|\bar p|^5} \nonumber\\
& &+\lambda E_\alpha \bar \nabla_\beta
\frac{2\bar p^\alpha \bar p^{\beta}{\bar p}^\mu + \bar p^2{\bar p}^\mu \Delta^{\alpha\beta}
-\bar p^2 \bar p^\alpha \Delta^{\mu\beta} -2\bar p^2 \bar p^\beta \Delta^{\mu\alpha} }{8|p|^6}
 \nonumber\\
& &+  B_\alpha \bar \nabla_\beta\frac{\bar \epsilon^{\beta\alpha\nu} \bar p_\nu \bar p^\mu + 2 \bar\epsilon^{\mu\beta\nu}\bar p_\nu \bar p^\alpha }{8|\bar p|^5}.
\end{eqnarray}
As we mentioned in previous sections, these quantities should be regarded as an operator instead of usual functions because the operator $\nabla$ also acts on the distribution function
The  specific expression for the CKE up to the second order is given by
\begin{eqnarray}
\label{CKE-full}
0&=& \dot{x}^{(0)\mu}\nabla_\mu \mathcal{J}_{n}
-  \frac{s B\cdot \bar p}{2|\bar p|^3} \hat{ \bar p}^\mu \bar \nabla_\mu \mathcal{J}_{n}
- \frac{\lambda s}{2|\bar p|^3}\bar\epsilon^{\mu\rho\sigma} E_\rho  {\bar p}_\sigma
\bar \nabla_\mu \mathcal{J}_{n}\nonumber\\
& & +\frac{3(B\cdot \bar p)^2 -\bar p^2 B^2 - 5(E\cdot \bar p)^2 - \bar p^2 E^2
 - 4\lambda |\bar p| \bar\epsilon^{\nu\rho\sigma}{\bar p}_\nu E_\rho B_\sigma }{8|\bar p|^6}
  \lambda \hat{ \bar p}^\mu\bar\nabla_\mu  \mathcal{J}_{n}\nonumber\\
& &+ \frac{(B\cdot \bar p)}{4|\bar p|^6} \bar\epsilon^{\mu\rho\sigma} E_\rho {\bar p}_\sigma  \bar \nabla_\mu  \mathcal{J}_{n}
 - \frac{3(E\cdot \bar p)  }{4|\bar p|^5}\lambda E^\mu \bar\nabla_\mu \mathcal{J}_{n}\nonumber\\
& & -\frac{ \bar p_\rho \bar p_{\sigma} -\bar p^2 \Delta_{\rho\sigma}}{8|\bar p|^4}\lambda\hat{ \bar p}^\mu\bar\nabla_\mu
\bar\nabla^\rho \bar \nabla^\sigma \mathcal{J}_{n}
+\frac{\bar \epsilon^{\nu\rho\sigma} \bar p_\rho  B_\sigma }{4|\bar p|^4}
 \lambda\hat{ \bar p}^\mu\bar\nabla_\mu\bar \nabla_\nu \mathcal{J}_{n}\nonumber\\
& &+\frac{2(E\cdot\bar p)(\bar p^2 \Delta^{\mu\nu} - \bar p^{\mu} \bar p^{\nu} )
+\bar p^2 \left[\bar p^\mu E^{\nu} - (E\cdot \bar p) \Delta^{\mu\nu}\right]}{4|\bar p|^6}
\bar\nabla_\mu \bar \nabla_\nu  \mathcal{J}_{n}.
\end{eqnarray}

We have organized the CKT in a different form  from the conventional one given in  \cite{Gao:2012ix,Stephanov:2012ki,Son:2012zy,Chen:2012ca,Manuel:2013zaa}.
This form is more convenient to be derived from Wigner function formalism and can be generalized to any higher order. If we drop all the second-order terms in the CKE above (\ref{CKE-full}),
only the first line remains. The remained first-order CKE can be verified to be  equivalent to the usual form \cite{Gao:2012ix,Stephanov:2012ki,Son:2012zy,Chen:2012ca,Manuel:2013zaa} up to first-order approximation, but in a more simple and elegant way.  All the terms associated with $\nabla $ in  Eqs.(\ref{omega-full}), (\ref{epsilon-full}) and (\ref{velocity-full}) and the terms including two or three operators $\nabla$
in (\ref{CKE-full}) are new and  have not been obtained in Ref.\cite{Gorbar:2017cwv}.
It should be useful to display  the results for $\sqrt{\omega}$ and $\varepsilon$ given in Ref.\cite{Gorbar:2017cwv} with our symbol conventions:
\begin{eqnarray}
\label{omega-full-I}
\sqrt{\omega} &=&1-\frac{s\lambda(B\cdot \bar p)}{2|\bar p|^3}
+\frac{1}{4|\bar p|^6}\left[2(B\cdot p)^2 - \bar p^2 B^2\right]
- \frac{\lambda}{2|\bar p|^5}\bar\epsilon^{\rho\mu\nu}\bar p_\rho E_\mu B_\nu, \\
\label{epsilon-full-I}
\varepsilon &=&|\bar p|+\frac{s\lambda(B\cdot \bar p)}{2|\bar p|^2}
+ \frac{1}{{16|\bar p|^5}} \left[-(B\cdot \bar p)^2 + 2B^2 \bar p^2  \right]
+\frac{\lambda}{4|\bar p|^4} \bar \epsilon^{\rho\mu\nu}  \bar p_\rho E_\mu B_\nu.
\end{eqnarray}
We note that the zeroth- and first-order expressions  are the same as our results, corresponding to the first and second terms on the right-handed sides of both equations, respectively.
The terms associated with $\bar \epsilon^{\rho\mu\nu}  \bar p_\rho E_\mu B_\nu$ are also the same. For $\sqrt{\omega}$, the third term on the right-handed side of the Eq.(\ref{omega-full-I})  is absent in
 our result (\ref{omega-full}).  Such difference can be eliminated by redefining  second-order distribution via the replacement in  (\ref{omega-full})
\begin{eqnarray}
{\mathcal{J}}^{(2)}_{sn} &\rightarrow&{\mathcal{J}}^{(2)}_{sn} - \frac{1}{4|\bar p|^6}\left[2(B\cdot p)^2 - \bar p^2 B^2\right]\mathcal{J}^{(0)}_{sn} .
\end{eqnarray}
Such replacement does not change the result (\ref{epsilon-full}). We note that the term associated with $B^2$ is the same between (\ref{epsilon-full}) and (\ref{epsilon-full-I}) while the term associated with
$(B\cdot p)^2$  is different. Another difference is that our result (\ref{epsilon-full}) includes the terms $E^2$ and $(E\cdot p)^2$ but they are  absent in  (\ref{epsilon-full-I}).
We interpret these differences as well as the addition terms with $\nabla$ in our result as the quantum effects from quantum field theory, which might not be able to be captured from semiclassical approach in quantum mechanics.
The difference between  our result and the one in \cite{Hayata:2020sqz,Mameda:2023ueq} is obvious. Our result here has already been written in seven-dimensional phase space and the Dirac delta functions or their
derivatives have been integrated out.  The results (\ref{omega-full}),(\ref{epsilon-full}),(\ref{velocity-full}), and (\ref{CKE-full})  can be used for numerical calculation directly and constitutes the  main result of  this paper.

\section{Summary}
\label{sec:summary}
Within the Wigner function formalism, we have derived the CKE in seven-dimensional phase space up to the second order in semiclassical expansion $\hbar$.
We find that second-order CKT includes various quantum effects which cannot be captured by the conventional semiclassical approach, such as the higher-derivative terms in coordinate space or momentum space. While this derived equation can be  the starting point for numerical evaluation on the transport process of non-linear chiral effects associated with  space-time derivative and electromagnetic fields, this paper also provides a systematic and consistent approach to derive the CKE  in seven-dimensional phase space from the original covariant  Wigner equations in eight-dimensional phase space
order by order. In the  present work, we restrict ourselves to the constant electromagnetic field in flat space-time. The generalization from constant to varying electromagnetic field, from massless to
massive fermions, from Abelian to non-Abelian gauge field, or from flat to curved space-time is straightforward and  will be investigated in the future.

\appendix

\section{Some calculation details  at second order}
\label{identities}

In this appendix, we will give some details on calculating the integral in Eq(\ref{Js-2-g-7d}). Let us start from Eq.(\ref{Jn-2-0})
by moving   the factor $1/p^2$ in the last term to be adjacent to $ \delta\left(p^2\right)$ and moving $\bar p_\mu$ and $\bar p_\rho$ to the left of the operator $\nabla_\rho $ and $\nabla_\mu $, respectively.
After such manipulation, we obtain
\begin{eqnarray}
\mathscr{J}^{(2)}_{n}
&=&  p_n  \tilde{\mathcal{J}}^{(2)}_{n} \delta\left(p^2\right)
+ \frac{s B\cdot p}{p^2} \left[  \tilde{\mathcal{J}}^{(1)}_{n} \delta\left(p^2\right) + \frac{s B\cdot p}{p_n}
\mathcal{J}^{(0)}_{n}\frac{ \delta\left(p^2\right)}{p^2}\right]\nonumber\\
& &+\frac{\bar\epsilon^{\rho\mu\nu}p_\rho E_\mu B_\nu}{2 p_n^2}\mathcal{J}^{(0)}_{n}\frac{ \delta\left(p^2\right)}{p^2}
\ {-}\frac{1}{2}\bar\epsilon_{\rho\mu\nu}B^\nu
\left(\nabla^\mu - \frac{2}{p^2}F^{\mu\sigma}p_\sigma\right)
\left[\frac{\bar p^{\rho}}{p_n} \mathcal{J}^{(0)}_{n} \frac{\delta\left(p^2\right)}{p^2} \right]\nonumber\\
& &{+}\frac{1}{4} \left(\bar\nabla_\rho-\frac{2}{p^2}F_{\rho\sigma}p^\sigma\right)
\left[\bar \epsilon^{\rho\mu\nu}B_\nu \frac{p_\mu}{p_n}  \mathcal{J}^{(0)}_{n} \frac{\delta\left(p^2\right)}{p^2} \right] \nonumber\\
& &\ {-}\frac{1}{4}\left( \bar\nabla_\rho-\frac{2}{p^2}F_{\rho\sigma}p^\sigma\right)
\left[ \frac{\bar p_\mu }{p_n^2} \left(E^\mu\bar p^{\rho} - E^\rho \bar p^{\mu}\right) \mathcal{J}^{(0)}_{n}\frac{ \delta\left(p^2\right)}{p^2} \right]\nonumber\\
& &+\frac{1}{4} \bar\nabla^\mu \bar \nabla^\nu
\left[ \frac{\bar p_\mu \bar p_{\nu} -\bar p^2 \Delta_{\mu\nu}}{p_n} \mathcal{J}^{(0)}_{n} \frac{\delta\left(p^2\right)}{p^2} \right]
\nonumber\\
& &-\frac{1}{2} \left( F^{\nu\rho}\nabla^\mu\frac{p_\rho}{p^2} + F^{\mu \rho}\nabla^\nu \frac{p_\rho}{p^2} \right)
\left[ \frac{\bar p_\mu \bar p_{\nu} -\bar p^2 \Delta_{\mu\nu}}{p_n} \mathcal{J}^{(0)}_{n} \frac{\delta\left(p^2\right)}{p^2} \right]
\nonumber\\
& &-\frac{1}{2} F^{\mu\rho}{F^{\nu\sigma}}\left( \frac{g_{\rho\sigma}}{p^2} -4  \frac{ p_\rho p_\sigma}{p^4}\right)
\left[ \frac{\bar p_\mu \bar p_{\nu} -\bar p^2 \Delta_{\mu\nu}}{p_n} \mathcal{J}^{(0)}_{n} \frac{\delta\left(p^2\right)}{p^2} \right].
\end{eqnarray}
where the operator $\nabla$ acts on all the  terms   following it.
Using  the relations
\begin{eqnarray}
 \delta'''\left(p^2\right)= - \frac{6\delta(p^2)}{p^6}, \ \ \  \delta''\left(p^2\right)=\frac{2\delta(p^2)}{p^4},\ \ \
 \delta'\left(p^2\right)=-\frac{\delta(p^2)}{p^2},
\end{eqnarray}
we can decompose $\mathscr{J}^{(2)}_{n}$ into seven parts
\begin{eqnarray}
\mathscr{J}^{(2)}_{n}
&=& \sum_{i=1}^7\mathscr{J}^{(2)}_{n,i}
\end{eqnarray}
where
\begin{eqnarray}
\mathscr{J}^{(2)}_{n,1}
&=& p_n  \tilde{\mathcal{J}}^{(2)}_{n} \delta\left(p^2\right)
+ \frac{s B\cdot p}{p^2} \left[  \tilde{\mathcal{J}}^{(1)}_{n} \delta\left(p^2\right) + \frac{s B\cdot p}{p_n}
\mathcal{J}^{(0)}_{n}\frac{ \delta\left(p^2\right)}{p^2}\right]\nonumber\\
\mathscr{J}^{(2)}_{n,2}
&=&-\frac{1}{2}\bar\epsilon^{\rho\mu\nu}\bar p_\rho E_\mu B_\nu\frac{1}{p_n^2}\mathcal{J}^{(0)}_{n} \delta'\left(p^2\right)\nonumber\\
\mathscr{J}^{(2)}_{n,3}&=& \frac{3}{4}\bar\epsilon_{\rho\mu\nu}B^\nu
\left[\bar \nabla^\mu \delta'(p^2) +\left(\bar\epsilon^{\mu\sigma\lambda}B_\sigma \bar p_\lambda+ E^{\mu}p_n\right) \delta''(p^2)\right]
\frac{\bar p^{\rho}}{p_n} \mathcal{J}^{(0)}_{n} \nonumber\\
\mathscr{J}^{(2)}_{n,4}&=&\frac{1}{4}\left[\bar \nabla_\rho \delta'(p^2)
+\left(\bar\epsilon_{\rho\sigma\nu}B^\sigma \bar p^\nu+ E_{\rho}p_n\right) \delta''(p^2)\right]
\frac{\bar p_\mu }{p_n^2} \left(E^\mu\bar p^{\rho} - E^\rho \bar p^{\mu}\right) \mathcal{J}^{(0)}_{n}\nonumber\\
\mathscr{J}^{(2)}_{n,5}&=&-\frac{1}{2} \left( \epsilon^{\mu\sigma\lambda}B_{\sigma}\nabla^\nu  \bar p_\lambda
 + E^{\mu}\nabla^\nu  p_n\right)
\frac{\bar p_\mu \bar p_{\nu} -\bar p^2 \Delta_{\mu\nu}}{p_n} \mathcal{J}^{(0)}_{n}\delta''\left(p^2\right)
\nonumber\\
\mathscr{J}^{(2)}_{n,6}&=&-\frac{1}{4}\Delta^\mu_\tau F^{\tau\rho} \Delta^\nu_\kappa F^{\kappa\sigma}
  \left[g_{\rho\sigma}\delta''(p^2) + \frac{4}{3}  p_\rho p_\sigma \delta'''(p^2)\right]
\frac{\bar p_\mu \bar p_{\nu} -\bar p^2 \Delta_{\mu\nu}}{p_n} \mathcal{J}^{(0)}_{n} \nonumber\\
\mathscr{J}^{(2)}_{n,7}&=&-\frac{1}{4} \bar\nabla^\mu \bar \nabla^\nu
\left[ \frac{\bar p_\mu \bar p_{\nu} -\bar p^2 \Delta_{\mu\nu}}{p_n} \mathcal{J}^{(0)}_{n} \delta'\left(p^2\right) \right]
\end{eqnarray}
Then we can finish integrating over $p_n$ by using the identities
\begin{eqnarray}
\delta\left(p^2\right)&=&\frac{1}{2|\bar p|}\sum_\lambda\delta\left(p_n-\lambda|\bar p|\right)\\
\delta'\left(p^2\right)&=&\frac{1}{4p_n|\bar p|}\sum_\lambda  \delta'\left(p_n-\lambda|\bar p|\right)\\
\delta''\left(p^2\right)&=&\frac{1}{8p_n|\bar p|^2}\sum_\lambda\left[\lambda \delta''\left(p_n -\lambda |\bar p|\right)
+\frac{1}{|\bar p|}\delta'\left(p_n-\lambda|\bar p|\right) \right]\\
\delta'''\left(p^2\right)&=&\frac{1}{16p_n |\bar p|^3}\sum_\lambda
\left[\delta'''\left(p_n-\lambda|\bar p|\right)
+\frac{3\lambda}{|\bar p|}\delta''\left(p_n-\lambda|\bar p|\right)
+\frac{3}{|\bar p|^2}\delta'\left(p_n-\lambda|\bar p|\right)\right]
\end{eqnarray}
The results after integration can be decomposed   into seven parts
\begin{eqnarray}
J_n^{(2)} & \equiv &  \int dp_n g(p_n) \mathscr{J}^{(2)}_{n} = \sum_{i=1}^7  J_{i}^{(2)},
\ \ \textrm{with}\ \  J_{i}^{(2)} \equiv \int dp_n g(p_n) \mathscr{J}^{(2)}_{n,i}
\end{eqnarray}
where
\begin{eqnarray}
 J_{1}^{(2)}&=& \int dp_n g(p_n)\left\{ p_n  \tilde{\mathcal{J}}^{(2)}_{n} \delta\left(p^2\right)
+ s B\cdot p\left[- \tilde{ \mathcal{J}}^{(1)}_{n} \delta'\left(p^2\right) + \frac{s B\cdot p}{2p_n}
\mathcal{J}^{(0)}_{n}\delta''\left(p^2\right)\right]\right\}\nonumber\\
&=&\sum_\lambda\left\{\frac{g}{2} \left[\lambda\tilde{ \mathcal{J}}^{(2)}_{n}
+\frac{s B\cdot \bar p}{2|\bar p|^2}\lambda\tilde{ \mathcal{J}}^{(1)\prime}_{n}
-  \frac{s B\cdot \bar p}{2|\bar p|^3} \left(\tilde{\mathcal{J}}^{(1)}_{n}
+ \frac{s B\cdot \bar p}{2|\bar p|^2} \mathcal{J}^{(0)\prime}_{n}\right) \right]\right.\nonumber\\
& &+\frac{g}{16 |\bar p|^2} \left( B\cdot \bar p\right)^2 \left(\lambda \mathcal{J}^{(0)\prime\prime}_{n}
-\frac{3}{|\bar p|} \mathcal{J}^{(0)\prime}_{n}
+ \frac{8\lambda}{|\bar p|^2} \mathcal{J}^{(0)}_{n}\right)\nonumber\\
& &+\frac{g'}{4|\bar p|^2} \left(s B\cdot \bar p\right)
\left(\lambda \tilde{\mathcal{J}}^{(1)}_{n}+ \frac{s B\cdot \bar p}{2|\bar p|^2}\lambda \mathcal{J}^{(0)\prime}_{n}
-\frac{5}{4}\frac{s B\cdot \bar p}{|\bar p|^3} \mathcal{J}^{(0)}_{n}\right)\nonumber\\
& &\left.+\frac{1}{16}g'' \frac{(B\cdot \bar p)^2}{|\bar p|^4}\lambda \mathcal{J}^{(0)}_{n}\right\}
\end{eqnarray}

\begin{eqnarray}
 J_{2}^{(2)}&=&-\frac{1}{2}\int dp_n g(p_n)
 \bar\epsilon^{\rho\mu\nu}\bar p_\rho E_\mu B_\nu\frac{1}{p_n^2}\mathcal{J}^{(0)}_{n} \delta'\left(p^2\right)\nonumber\\
&=&\sum_\lambda\left\{ \frac{g}{8|\bar p|^4}\bar\epsilon^{\rho\mu\nu}\bar p_\rho E_\mu B_\nu
\left(\lambda  \mathcal{J}^{(0)\prime}_{n}
- \frac{3}{|\bar p|} \mathcal{J}^{(0)}_{n} \right)
+\frac{g'}{8|\bar p|^4}\bar\epsilon^{\rho\mu\nu}\bar p_\rho E_\mu B_\nu
\lambda\mathcal{J}^{(0)}_{n}\right\}
\end{eqnarray}

\begin{eqnarray}
 J_{3}^{(2)}&=&\frac{3}{4}\int dp_n g(p_n)
 \bar\epsilon_{\rho\mu\nu}B^\nu
\left[\bar \nabla^\mu \delta'(p^2) +\left(\bar\epsilon^{\mu\sigma\lambda}B_\sigma \bar p_\lambda+ E^{\mu}p_n\right) \delta''(p^2)\right]
\frac{\bar p^{\rho}}{p_n} \mathcal{J}^{(0)}_{n}\nonumber\\
&=&\sum_\lambda \left\{ {-}\frac{3g}{16}  \bar\epsilon_{\rho\mu\nu}B^\nu
\bar \nabla^\mu  \frac{\bar p^{\rho}}{|\bar p|^3} \left( \mathcal{J}^{(0)\prime}_{n}
- \frac{ 2\lambda  }{|\bar p|} \mathcal{J}^{(0)}_{n} \right)\right.  \nonumber\\
& & {+}\frac{3g}{32|\bar p|^4}\left[(B\cdot \bar p)^2-B^2 \bar p^2\right]
\left(\lambda\mathcal{J}^{(0)\prime\prime}_{n}
 -\frac{5 }{|\bar p|} \mathcal{J}^{(0)\prime}_{n}
+\frac{8\lambda }{|\bar p|^2}\mathcal{J}^{(0)}_{n} \right)\nonumber\\
& & {+}\frac{3g}{32|\bar p|^3} \bar\epsilon_{\rho\mu\nu}\bar p^{\rho} E^{\mu} B^\nu
\left( \mathcal{J}^{(0)\prime\prime}_{n}
-\frac{3\lambda }{|\bar p|} \mathcal{J}^{(0)\prime}_{n}
+\frac{3}{|\bar p|^2} \mathcal{J}^{(0)}_{n} \right)\nonumber\\
& & {-}\frac{3g'}{16|\bar p|^3}  \bar\epsilon_{\rho\mu\nu}\bar p^{\rho} B^\nu
\bar \nabla^\mu \mathcal{J}^{(0)}_{n}
{+}\frac{ g'}{8|\bar p|^5} \left[3 (B\cdot \bar p)^2 - B^2 \bar p^2\right] \mathcal{J}^{(0)}_{n}\nonumber\\
& & {-}\frac{3}{32|\bar p|^4}\left(\frac{ 1}{|\bar p|}g'  +\lambda g''\right)
\left[(B\cdot \bar p)^2-B^2 \bar p^2\right]
\mathcal{J}^{(0)}_{n}\nonumber\\
& & \left. {+}\frac{3}{32|\bar p|^3}\left(\frac{\lambda }{|\bar p|}g' -g''\right)
 \bar\epsilon_{\rho\mu\nu}\bar p^{\rho} E^{\mu} B^\nu\mathcal{J}^{(0)}_{n}\right\}
\end{eqnarray}

\begin{eqnarray}
J_{4}^{(2)}&=&\frac{1}{4}\int dp_n g(p_n)\left[\bar \nabla_\rho \delta'(p^2)
+\left(\bar\epsilon_{\rho\sigma\lambda}B^\sigma \bar p^\lambda+ E_{\rho}p_n\right) \delta''(p^2)\right]
\frac{\bar p_\mu }{p_n^2} \left(E^\mu\bar p^{\rho} - E^\rho \bar p^{\mu}\right) \mathcal{J}^{(0)}_{n} \nonumber\\
&=&\sum_\lambda\left\{ {-}\frac{g }{16}\bar \nabla_\rho \frac{(E\cdot \bar p)\bar p^{\rho} - E^\rho \bar p^2}{|\bar p|^4}
 \left(\lambda  \mathcal{J}^{(0)\prime}_{n} -\frac{3}{|\bar p|} \mathcal{J}^{(0)}_{n}  \right)\right.\nonumber\\
& & {+}\frac{g}{32|\bar p|^4} \left[(E\cdot \bar p)^2  - E^2 \bar p^2\right]
\left(\lambda \mathcal{J}^{(0)\prime\prime}_{n}
- \frac{5  }{|\bar p|} \mathcal{J}^{(0)\prime}_{n}
+ \frac{8\lambda  }{|\bar p|^2} \mathcal{J}^{(0)}_{n} \right)\nonumber\\
& & {+}\frac{g}{32|\bar p|^3}\bar\epsilon_{\rho \mu\nu} \bar p^\rho E^\mu B^\nu
\left( \mathcal{J}^{(0)\prime\prime}_{n}
- \frac{7\lambda }{|\bar p|} \mathcal{J}^{(0)\prime}_{n}
+ \frac{15  }{|\bar p|^2} \mathcal{J}^{(0)}_{n} \right)\nonumber\\
& & {-}\frac{\lambda g'}{16|\bar p|^4}  \left[(E\cdot \bar p)\bar p^{\rho} - E^\rho \bar p^2\right]
\bar \nabla_\rho \mathcal{J}^{(0)}_{n} \nonumber\\
& & {+}\frac{1}{32|\bar p|^4}\left(  \frac{ 1 }{|\bar p|}g'-\lambda g'' \right)
\left[(E\cdot \bar p)^2  - E^2 \bar p^2\right]\mathcal{J}^{(0)}_{n}\nonumber\\
& & \left. {+}\frac{1}{32|\bar p|^3}\left( \frac{\lambda  }{|\bar p|}g'- g''\right)
\bar\epsilon_{\rho \mu\nu} \bar p^\rho E^\mu B^\nu\mathcal{J}^{(0)}_{n}\right\}
\end{eqnarray}

\begin{eqnarray}
\label{J-2-5}
J_{5}^{(2)}&=&-\frac{1}{2}\int dp_n g(p_n) \left(\bar \epsilon^{\mu\sigma\lambda}B_{\sigma}\bar\nabla^\nu  \bar p_\lambda
+ E^{\mu}\bar\nabla^\nu  p_n\right)
\frac{\bar p_\mu \bar p_{\nu} -\bar p^2 \Delta_{\mu\nu}}{p_n} \mathcal{J}^{(0)}_{n}\delta''\left(p^2\right) \nonumber\\
 &=&\sum_\lambda\left\{-\frac{g}{16}\bar \epsilon^{\mu\sigma\lambda}B_{\sigma} \bar\nabla_\mu \frac{ \bar p_\lambda}{|\bar p|^2}
 \left(\lambda\mathcal{J}^{(0)\prime\prime}_{n}
 -\frac{5}{|\bar p|}\mathcal{J}^{(0)\prime}_{n}
 + \frac{8\lambda }{|\bar p|^2}\mathcal{J}^{(0)}_{n}\right)\right.\nonumber\\
& &-\frac{g}{16} E^{\mu} \bar \nabla^\nu
\frac{\bar p_\mu \bar p_{\nu} -\bar p^2 \Delta_{\mu\nu}}{|\bar p|^3}
\left(\mathcal{J}^{(0)\prime\prime}_{n}
-\frac{3\lambda }{|\bar p|}\mathcal{J}^{(0)\prime}_{n}
+\frac{3 }{|\bar p|^2}\mathcal{J}^{(0)}_{n}\right)\nonumber\\
& & +\frac{1}{16|\bar p|^2}\left(\frac{5}{|\bar p|}g'-\lambda g''\right)
\bar \epsilon^{\mu\sigma\lambda}B_{\sigma}   \bar p_\lambda
\bar\nabla_\mu \mathcal{J}^{(0)}_{n} \nonumber\\
& & +\frac{5g'}{16|\bar p|^5}\left[B^2\bar p^2 - 3(B\cdot \bar p)^2\right]\mathcal{J}^{(0)}_{n}
+\frac{\lambda g''}{8|\bar p|^4} (B\cdot \bar p)^2 \mathcal{J}^{(0)}_{n}\nonumber\\
& &+\frac{1}{16|\bar p|^3}\left(\frac{3\lambda }{|\bar p|}g' - g'' \right)
\left[(E\cdot \bar p) \bar p_{\nu} -\bar p^2 E_{\nu}\right)
\bar \nabla^\nu\mathcal{J}^{(0)}_{n}\nonumber\\
& &-\frac{1}{16|\bar p|^3} \left(\frac{8 }{|\bar p|^2}g'
 -\frac{5 \lambda }{|\bar p|}g'' +  g'''\right)
\left[(B\cdot p)^2 - B^2 \bar p^2\right]
\mathcal{J}^{(0)}_{n}\nonumber\\
& &-\frac{1}{16|\bar p|^2}\left(\frac{14\lambda }{|\bar p|^2}g'-\frac{8 }{|\bar p|}g''+2 \lambda g'''\right)
\bar\epsilon^{\rho\mu\nu}\bar p_\rho E_\mu B_\nu
\mathcal{J}^{(0)}_{n}\nonumber\\
& &-\frac{1}{16|\bar p|^3}\left(\frac{3 }{|\bar p|^2}g' -\frac{3\lambda }{|\bar p|}g''+ g'''\right)
\left[(E\cdot \bar p)^2-E^2 \bar p^2\right]
\mathcal{J}^{(0)}_{n} \nonumber\\
& &{-\frac{\lambda g'}{8|\bar p|^2}\bar \epsilon^{\mu\rho\sigma}B_{\rho}   \bar p_\sigma
\bar\nabla_\mu \mathcal{J}^{(0)\prime}_{n}
 +\frac{\lambda g'}{4|\bar p|^4}(B\cdot \bar p)^2  \mathcal{J}^{(0)\prime}_{n} }\nonumber\\
& &{-\frac{g'}{8|\bar p|^3}\left[ (E\cdot \bar p)\bar p_{\nu} - \bar p^2 E_\nu \right]
\bar \nabla^\nu \mathcal{J}^{(0)\prime}_{n}}\nonumber\\
& &{-\frac{1}{16|\bar p|^3}\left[(B\cdot p)^2 - B^2 \bar p^2\right]
 \left(g' \mathcal{J}^{(0)\prime\prime}_{n}
 -\frac{5\lambda g'}{|\bar p|}\mathcal{J}^{(0)\prime}_{n}
 +2 g'' \mathcal{J}^{(0)\prime}_{n} \right)}\nonumber\\
& &{-\frac{1}{16|\bar p|^2}\bar\epsilon^{\rho\mu\nu}\bar p_\rho E_\mu B_\nu
\left(2\lambda g'\mathcal{J}^{(0)\prime\prime}_{n}
-\frac{8 g'}{|\bar p|}\mathcal{J}^{(0)\prime}_{n}
+4\lambda g''\mathcal{J}^{(0)\prime}_{n}\right)}\nonumber\\
& &\left.{-\frac{1}{16|\bar p|^3}\left[(E\cdot \bar p)^2-E^2 \bar p^2\right]
\left(g'\mathcal{J}^{(0)\prime\prime}_{n}
-\frac{3\lambda g'}{|\bar p|}\mathcal{J}^{(0)\prime}_{n}
+2g''\mathcal{J}^{(0)\prime}_{n}\right)}\right\}
\end{eqnarray}

\begin{eqnarray}
\label{J-2-6}
J_6^{(2)} &=& -\frac{1}{4}\int dp_n g(p_n)\Delta^\mu_\lambda F^{\lambda\rho} \Delta^\nu_\kappa F^{\kappa\sigma}
 \left[g_{\rho\sigma}\delta''(p^2) + \frac{4}{3}  p_\rho p_\sigma \delta'''(p^2)\right]
\frac{\bar p_\mu \bar p_{\nu} -\bar p^2 \Delta_{\mu\nu}}{p_n} \mathcal{J}^{(0)}_{n}\nonumber\\
&=&\sum_\lambda\left\{-\frac{g}{32|\bar p|^4}\left[(E\cdot\bar p)^2 -E^2 \bar p^2  + (B\cdot \bar p)^2 + B^2 \bar p^2\right]
\left(\lambda\mathcal{J}^{(0)\prime\prime}_{n}
-\frac{5}{|\bar p|}\mathcal{J}^{(0)\prime}_{n}
 +\frac{8\lambda }{|\bar p|^2}\mathcal{J}^{(0)}_{n} \right)\right.\nonumber\\
& &+\frac{g}{48|\bar p|^3}\left[(B\cdot \bar p)^2-B^2 \bar p^2\right]
\left(\mathcal{J}^{(0)\prime\prime\prime}_{n}
-\frac{9\lambda }{|\bar p|}\mathcal{J}^{(0)\prime\prime}_{n}
+ \frac{33 }{|\bar p|^2}\mathcal{J}^{(0)\prime}_{n}
-\frac{48\lambda }{|\bar p|^6}\mathcal{J}^{(0)}_{n}\right)\nonumber\\
& &+\frac{g}{24|\bar p|^2}\bar\epsilon^{\rho\mu\nu}\bar p_\rho E_\mu B_\nu
\left(\lambda \mathcal{J}^{(0)\prime\prime\prime}_{n}
-\frac{6}{|\bar p|}\mathcal{J}^{(0)\prime\prime}_{n}
+ \frac{15\lambda}{|\bar p|^2}\mathcal{J}^{(0)\prime}_{n}
-\frac{15 }{|\bar p|^3}\mathcal{J}^{(0)}_{n}\right)\nonumber\\
& &+\frac{g}{48|\bar p|^3}\left[(E\cdot\bar p)^2 -E^2 \bar p^2 \right]
\left(\mathcal{J}^{(0)\prime\prime\prime}_{n}
-\frac{3\lambda }{|\bar p| }\mathcal{J}^{(0)\prime\prime}_{n}
+\frac{3 }{|\bar p|^2 }\mathcal{J}^{(0)\prime}_{n}\right)\nonumber\\
& &+\frac{1}{32|\bar p|^4}\left(\frac{5}{|\bar p|}g' -\lambda g'' \right)
\left[(E\cdot\bar p)^2 -E^2 \bar p^2 + (B\cdot \bar p)^2 + B^2 \bar p^2\right]
\mathcal{J}^{(0)}_{n}\nonumber\\
& &+\frac{1}{48|\bar p|^3}\left(\frac{33 }{|\bar p|^2}g'
-\frac{9\lambda }{|\bar p|}g''+ g'''\right)
\left[(B\cdot \bar p)^2-B^2 \bar p^2\right]
\mathcal{J}^{(0)}_{n} \nonumber\\
& &+\frac{1}{48|\bar p|^3}\left( \frac{3}{|\bar p|^2}g' - \frac{3\lambda }{|\bar p|}g''+ g''' \right)
\left[(E\cdot\bar p)^2 -E^2 \bar p^2 \right]
\mathcal{J}^{(0)}_{n}\nonumber\\
& &+\frac{1}{24|\bar p|^2}\left( \frac{15\lambda }{|\bar p|^2}g'
-\frac{6}{|\bar p|}g''+ \lambda g'''\right)
\bar\epsilon^{\rho\mu\nu}\bar p_\rho E_\mu B_\nu
\mathcal{J}^{(0)}_{n}\nonumber\\
& &{-\frac{\lambda g'}{16|\bar p|^4}\left[(E\cdot\bar p)^2 -E^2 \bar p^2 + (B\cdot \bar p)^2 + B^2 \bar p^2\right]
\mathcal{J}^{(0)\prime}_{n}}\nonumber\\
& &{+\frac{1}{16|\bar p|^3}\left[(B\cdot \bar p)^2-B^2 \bar p^2\right]
\left(g'\mathcal{J}^{(0)\prime\prime}_{n}
-\frac{6\lambda }{|\bar p|}g'\mathcal{J}^{(0)\prime}_{n}
+ g''\mathcal{J}^{(0)\prime}_{n}\right)}\nonumber\\
& &{+\frac{1}{16|\bar p|^3}\left[(E\cdot\bar p)^2 -E^2 \bar p^2 \right]
\left(  g'\mathcal{J}^{(0)\prime\prime}
-\frac{2\lambda }{|\bar p|}g'\mathcal{J}^{(0)\prime}
+ g''\mathcal{J}^{(0)\prime}  \right)}\nonumber\\
& &\left.{+\frac{1}{8|\bar p|^2}\bar\epsilon^{\rho\mu\nu}\bar p_\rho E_\mu B_\nu
\left(\lambda g'\mathcal{J}^{(0)\prime\prime}_{n}
-\frac{4}{|\bar p|}g'\mathcal{J}^{(0)\prime}_{n}
+\lambda g''\mathcal{J}^{(0)\prime}_{n}\right)}\right\}
\end{eqnarray}

\begin{eqnarray}
\label{J-2-7}
J_7^{(2)}&=& -\frac{1}{4}\int dp_n g(p_n) \bar\nabla^\mu \bar \nabla^\nu
\left[ \frac{\bar p_\mu \bar p_{\nu} -\bar p^2 \Delta_{\mu\nu}}{p_n} \mathcal{J}^{(0)}_{n} \delta'\left(p^2\right) \right]\nonumber\\
 &=&\sum_\lambda\left\{ \frac{g}{16}  \bar\nabla^\mu \bar \nabla^\nu \frac{\bar p_\mu \bar p_{\nu} -\bar p^2 \Delta_{\mu\nu}}{|\bar p|^3}
\left( \mathcal{J}^{(0)\prime}_{n}-\frac{2\lambda }{|\bar p|}\mathcal{J}^{(0)}_{n}  \right)\right.\nonumber\\
& & + \frac{g' }{16}   \left(\bar p_\mu \bar p_{\nu} -\bar p^2 \Delta_{\mu\nu}\right)
\frac{1}{|\bar p|^3}\bar\nabla^\mu \bar \nabla^\nu \mathcal{J}^{(0)}_{n} \nonumber\\
& &-\frac{1}{8|\bar p|^2}\left( \frac{2  }{|\bar p|}g' - \lambda g''\right)\bar \epsilon^{\mu\rho\sigma} B_\rho \bar p_\sigma
\bar \nabla_\mu \mathcal{J}^{(0)}_{n}  \nonumber\\
& &-\frac{1}{8|\bar p|^3}\left(\frac{2\lambda }{|\bar p|}g' -  g'' \right)
\left[ (E\cdot\bar p) \bar p_{\nu} -\bar p^2 E_{\nu}\right]
\bar \nabla^\nu\mathcal{J}^{(0)}_{n}  \nonumber\\
& & + \frac{g'}{4|\bar p|^5} \left[2(B\cdot\bar p)^2 - B^2\bar p^2 \right] \mathcal{J}^{(0)}_{n}
 - \frac{\lambda g''}{8|\bar p|^4 }  (B\cdot \bar p)^2  \mathcal{J}^{(0)}_{n} \nonumber\\
& & - \frac{1}{16|\bar p|^3} \left(\frac{2 \lambda }{|\bar p|}g''-  g'''  \right)
\left[(B\cdot \bar p)^2 -B^2 \bar p^2\right]\mathcal{J}^{(0)}_{n}\nonumber\\
& &-\frac{1}{16|\bar p|^3}\left( \frac{2\lambda }{|\bar p|}g'' -  g''' \right)
\left[(E\cdot \bar p)^2 -\bar p^2 E^2\right]\mathcal{J}^{(0)}_{n} \nonumber\\
& &+\frac{1}{8|\bar p|^2} \left( \frac{2\lambda }{|\bar p|^2}g' - \frac{2 }{|\bar p|}g''
+ \lambda g'''  \right)
\bar \epsilon^{\mu\rho\sigma} E_\mu B_\rho \bar p_\sigma
\mathcal{J}^{(0)}_{n}\nonumber\\
& &{+ \frac{\lambda g'}{8|\bar p|^2}   \bar\epsilon^{\mu \rho\sigma} B_\rho \bar p_\sigma
\bar\nabla_\mu \mathcal{J}^{(0)\prime}_{n}
+\frac{g'}{8|\bar p|^3} E^\nu \left(\bar p_\mu \bar p_{\nu} -\bar p^2 \Delta_{\mu\nu}\right)
\bar\nabla^\mu \mathcal{J}^{(0)\prime}_{n}}\nonumber\\
& &{ - \frac{\lambda g'}{8|\bar p|^4}  (B\cdot \bar p)^2 \mathcal{J}^{(0)\prime}_{n}
+ \frac{ g''}{16|\bar p|^3}  \left[(B\cdot \bar p)^2 -B^2 \bar p^2\right]\mathcal{J}^{(0)\prime}_{n} }\nonumber\\
& &\left. {+\frac{g''}{16|\bar p|^3}\left[(E\cdot \bar p)^2 -\bar p^2 E^2\right]\mathcal{J}^{(0)\prime}_{n}
+\frac{\lambda g''}{8|\bar p|^2} \bar \epsilon^{\mu\rho\sigma} E_\mu B_\rho \bar p_\sigma \mathcal{J}^{(0)\prime}_{n}  }\right\}
\end{eqnarray}
After putting all the $J_i^{(2)}$ together, we arrive at  Eq.(\ref{Js-2-g-7d}).  We find that the terms associated with the mixed derivative terms of $g$ and $\mathcal{J}^{(0)}_{n}$  including the last five lines of $J_5^{(2)}$ in Eq.(\ref{J-2-5}), the last four lines of $J_6^{(2)}$ in Eq.(\ref{J-2-6}),
 and the last three lines of $J_7^{(2)}$ in Eq.(\ref{J-2-7}) cancel each other. Such cancelation makes the final physical quantities only depend on
 the distribution function $\mathcal{J}^{(2)}_{n}$,  $\mathcal{J}^{(1)}_{n}$, and $\mathcal{J}^{(0)}_{n}$.

\acknowledgments

This work was supported in part by  the National Natural
Science Foundation of China  under Grant
Nos. 12175123, 12321005




\begin{thebibliography}{99}








\bibitem{Bzdak:2011yy}
  A.~Bzdak and V.~Skokov,
  Phys.\ Lett.\ B {\bf 710} (2012) 171

\bibitem{Deng:2012pc}
  W.~T.~Deng and X.~G.~Huang,
  Phys.\ Rev.\ C {\bf 85} (2012) 044907

\bibitem{Bloczynski:2012en}
  J.~Bloczynski, X.~G.~Huang, X.~Zhang and J.~Liao,
  Phys.\ Lett.\ B {\bf 718} (2013) 1529


\bibitem{Liang:2004ph}
  Z.~T.~Liang and X.~N.~Wang,
  Phys.\ Rev.\ Lett.\  {\bf 94} (2005) 102301
  Erratum: [Phys.\ Rev.\ Lett.\  {\bf 96} (2006) 039901]

\bibitem{Gao:2007bc}
  J.~H.~Gao, S.~W.~Chen, W.~t.~Deng, Z.~T.~Liang, Q.~Wang and X.~N.~Wang,
  Phys.\ Rev.\ C {\bf 77} (2008) 044902

\bibitem{Becattini:2007sr}
  F.~Becattini, F.~Piccinini and J.~Rizzo,
  Phys.\ Rev.\ C {\bf 77} (2008) 024906


\bibitem{Csernai:2013bqa}
  L.~P.~Csernai, V.~K.~Magas and D.~J.~Wang,
  Phys.\ Rev.\ C {\bf 87} (2013) no.3,  034906

Jiang:2016woz,Deng:2016gyh,Pang:2016igs
\bibitem{Jiang:2016woz}
  Y.~Jiang, Z.~W.~Lin and J.~Liao,
  Phys.\ Rev.\ C {\bf 94} (2016) no.4,  044910
   Erratum: [Phys.\ Rev.\ C {\bf 95} (2017) no.4,  049904]

\bibitem{Deng:2016gyh}
  W.~T.~Deng and X.~G.~Huang,
  Phys.\ Rev.\ C {\bf 93} (2016) no.6,  064907

\bibitem{Pang:2016igs}
  L.~G.~Pang, H.~Petersen, Q.~Wang and X.~N.~Wang,
  Phys.\ Rev.\ Lett.\  {\bf 117} (2016) no.19,  192301









\bibitem{Vilenkin:1980fu}
  A.~Vilenkin,
  Phys.\ Rev.\ D {\bf 22}, 3080 (1980).

\bibitem{Kharzeev:2007jp}
  D.~E.~Kharzeev, L.~D.~McLerran and H.~J.~Warringa,
  Nucl.\ Phys.\ A {\bf 803}, 227 (2008).

\bibitem{Fukushima:2008xe}
  K.~Fukushima, D.~E.~Kharzeev and H.~J.~Warringa,
  Phys.\ Rev.\ D {\bf 78}, 074033 (2008).


\bibitem{Vilenkin:1978hb}
  A.~Vilenkin,
  Phys.\ Lett.\  {\bf 80B}, 150 (1978).


\bibitem{Kharzeev:2007tn}
  D.~Kharzeev and A.~Zhitnitsky,
  Nucl.\ Phys.\ A {\bf 797} (2007) 67

\bibitem{Erdmenger:2008rm}
  J.~Erdmenger, M.~Haack, M.~Kaminski and A.~Yarom,
  JHEP {\bf 0901}, 055 (2009)

\bibitem{Banerjee:2008th}
  N.~Banerjee, J.~Bhattacharya, S.~Bhattacharyya, S.~Dutta, R.~Loganayagam and P.~Surowka,
  JHEP {\bf 1101}, 094 (2011)


\bibitem{Son:2004tq}
  D.~T.~Son and A.~R.~Zhitnitsky,
  Phys.\ Rev.\ D {\bf 70} (2004) 074018

\bibitem{Metlitski:2005pr}
  M.~A.~Metlitski and A.~R.~Zhitnitsky,
  Phys.\ Rev.\ D {\bf 72} (2005) 045011




\bibitem{Stephanov:2012ki}
  M.~A.~Stephanov and Y.~Yin,
  Phys.\ Rev.\ Lett.\  {\bf 109} (2012) 162001

\bibitem{Gao:2012ix}
  J.~H.~Gao, Z.~T.~Liang, S.~Pu, Q.~Wang and X.~N.~Wang,
  Phys.\ Rev.\ Lett.\  {\bf 109}, 232301 (2012)

\bibitem{Son:2012zy}
  D.~T.~Son and N.~Yamamoto,
  Phys.\ Rev.\ D {\bf 87} (2013) 085016


\bibitem{Chen:2012ca}
  J.~W.~Chen, S.~Pu, Q.~Wang and X.~N.~Wang,
  Phys.\ Rev.\ Lett.\  {\bf 110} (2013) no.26,  262301

\bibitem{Manuel:2013zaa}
  C.~Manuel and J.~M.~Torres-Rincon,
  Phys.\ Rev.\ D {\bf 89} (2014) no.9,  096002

\bibitem{Chen:2014cla}
  J.~Y.~Chen, D.~T.~Son, M.~A.~Stephanov, H.~U.~Yee and Y.~Yin,
  Phys.\ Rev.\ Lett.\  {\bf 113} (2014) no.18,  182302

\bibitem{Chen:2015gta}
  J.~Y.~Chen, D.~T.~Son and M.~A.~Stephanov,
  Phys.\ Rev.\ Lett.\  {\bf 115} (2015) no.2,  021601


\bibitem{Hidaka:2016yjf}
  Y.~Hidaka, S.~Pu and D.~L.~Yang,
  Phys.\ Rev.\ D {\bf 95} (2017) no.9,  091901

\bibitem{Mueller:2017lzw}
  N.~Mueller and R.~Venugopalan,
  Phys.\ Rev.\ D {\bf 97} (2018) no.5,  051901

\bibitem{Huang:2018wdl}
  A.~Huang, S.~Shi, Y.~Jiang, J.~Liao and P.~Zhuang,
  Phys.\ Rev.\ D {\bf 98} (2018) no.3,  036010

\bibitem{Hidaka:2018ekt}
  Y.~Hidaka and D.~L.~Yang,
  Phys.\ Rev.\ D {\bf 98} (2018) no.1,  016012

\bibitem{Gao:2018wmr}
  J.~H.~Gao, Z.~T.~Liang, Q.~Wang and X.~N.~Wang,
  Phys.\ Rev.\ D {\bf 98} (2018) no.3,  036019

\bibitem{Gao:2018jsi}
  J.~H.~Gao, J.~Y.~Pang and Q.~Wang,
  Phys.\ Rev.\ D {\bf 100} (2019) no.1,  016008


\bibitem{Liu:2018xip}
  Y.~C.~Liu, L.~L.~Gao, K.~Mameda and X.~G.~Huang,
  Phys.\ Rev.\ D {\bf 99} (2019) no.8,  085014


\bibitem{Liu:2020ymh}
Y.~C.~Liu and X.~G.~Huang,
Nucl. Sci. Tech. \textbf{31} (2020) no.6, 56

\bibitem{Gao:2020vbh}
J.~H.~Gao, G.~L.~Ma, S.~Pu and Q.~Wang,
Nucl. Sci. Tech. \textbf{31} (2020) no.9, 90

\bibitem{Gao:2020pfu}
J.~H.~Gao, Z.~T.~Liang and Q.~Wang,
Int. J. Mod. Phys. A \textbf{36}, no.01, 2130001 (2021)

\bibitem{Hidaka:2022dmn}
Y.~Hidaka, S.~Pu, Q.~Wang and D.~L.~Yang,
Prog. Part. Nucl. Phys. \textbf{127}, 103989 (2022)



\bibitem{Gorbar:2016ygi}
E.~V.~Gorbar, V.~A.~Miransky, I.~A.~Shovkovy and P.~O.~Sukhachov,
Phys. Rev. Lett. \textbf{118}, no.12, 127601 (2017)


\bibitem{Gorbar:2016sey}
E.~V.~Gorbar, V.~A.~Miransky, I.~A.~Shovkovy and P.~O.~Sukhachov,
Phys. Rev. B \textbf{95}, no.11, 115202 (2017)



\bibitem{Gorbar:2016vvg}
E.~V.~Gorbar, V.~A.~Miransky, I.~A.~Shovkovy and P.~O.~Sukhachov,
Phys. Rev. B \textbf{95}, no.11, 115422 (2017)





\bibitem{Yamamoto:2020zrs}
N.~Yamamoto and D.~L.~Yang,
Astrophys. J. \textbf{895}, no.1, 56 (2020)



\bibitem{Kamada:2022nyt}
K.~Kamada, N.~Yamamoto and D.~L.~Yang,
Prog. Part. Nucl. Phys. \textbf{129}, 104016 (2023)


\bibitem{Kharzeev:2011ds}
  D.~E.~Kharzeev and H.~-U.~Yee,
  Phys.\ Rev.\ D {\bf 84}, 045025 (2011).

\bibitem{Banerjee:2012iz}
  N.~Banerjee, J.~Bhattacharya, S.~Bhattacharyya, S.~Jain, S.~Minwalla and T.~Sharma,
  JHEP {\bf 1209} (2012) 046

\bibitem{Bhattacharyya:2013ida}
  S.~Bhattacharyya, J.~R.~David and S.~Thakur,
  JHEP {\bf 1401} (2014) 010

\bibitem{Megias:2014mba}
  E.~Megias and M.~Valle,
  JHEP {\bf 1411} (2014) 005

\bibitem{Jimenez-Alba:2015bia}
  A.~Jimenez-Alba and H.~U.~Yee,
  Phys.\ Rev.\ D {\bf 92} (2015) no.1,  014023

\bibitem{Satow:2014lia}
  D.~Satow,
  Phys.\ Rev.\ D {\bf 90} (2014) no.3,  034018

\bibitem{Gorbar:2017toh}
  E.~V.~Gorbar, D.~O.~Rybalka and I.~A.~Shovkovy,
  Phys.\ Rev.\ D {\bf 95} (2017) no.9,  096010

\bibitem{Abbasi:2018zoc}
  N.~Abbasi, F.~Taghinavaz and O.~Tavakol,
  JHEP {\bf 1903} (2019) 051





\bibitem{Gorbar:2017cwv}
  E.~V.~Gorbar, V.~A.~Miransky, I.~A.~Shovkovy and P.~O.~Sukhachov,
  Phys.\ Rev.\ B {\bf 95} (2017) no.20,  205141

\bibitem{Hayata:2020sqz}
T.~Hayata, Y.~Hidaka and K.~Mameda,
JHEP \textbf{05}, 023 (2021)

\bibitem{Mameda:2023ueq}
K.~Mameda,
Phys. Rev. D \textbf{108}, no.1, 016001 (2023)




\bibitem{Heinz:1983nx}
  U.~W.~Heinz,
  Phys.\ Rev.\ Lett.\  {\bf 51} (1983) 351.

\bibitem{Elze:1986qd}
  H.~T.~Elze, M.~Gyulassy and D.~Vasak,
  Nucl.\ Phys.\ B {\bf 276} (1986) 706.

\bibitem{Vasak:1987um}
  D.~Vasak, M.~Gyulassy and H.~T.~Elze,
  Annals Phys.(N.Y.)\  {\bf 173} (1987) 462.

\bibitem{Zhuang:1995pd}
  P.~Zhuang and U.~W.~Heinz,
  Annals Phys.\  {\bf 245} (1996) 311.

\bibitem{Gao:2015zka}
  J.~h.~Gao and Q.~Wang,
  Phys.\ Lett.\ B {\bf 749} (2015) 542

\bibitem{Gao:2017gfq}
  J.~h.~Gao, S.~Pu and Q.~Wang,
  Phys.\ Rev.\ D {\bf 96} (2017) no.1,  016002

\end{thebibliography}
\end{document}